\documentclass[11pt]{article}
\usepackage{emnlp2021}
\usepackage{times}
\usepackage{latexsym}
\usepackage[T1]{fontenc}
\usepackage[utf8]{inputenc}
\usepackage{microtype}
\usepackage{subcaption}
\usepackage{microtype}
\usepackage{multirow}
\usepackage{enumitem}
\usepackage{rotating}
\usepackage{amsfonts}
\usepackage{amsmath}
\usepackage{soul}
\usepackage{mathtools}
\usepackage{chngcntr}
\usepackage{graphicx}
\usepackage{soul,xcolor}
\usepackage{booktabs}

\newcommand{\firstdata}{\textsf{Harm-C}}
\newcommand{\seconddata}{\textsf{Harm-P}}
\newcommand{\acldata}{\textsf{HarMeme}}
\newcommand{\model}{\texttt{MOMENTA}}
\title{MOMENTA: A Multimodal Framework\\
for Detecting Harmful Memes and Their Targets}

\author{Shraman Pramanick$^1$\thanks{\;\;denotes equal contribution}, Shivam Sharma$^{2,4}$\footnotemark[1], Dimitar Dimitrov$^3$, Md Shad Akhtar$^2$, \\ \textbf{Preslav Nakov$^5$, Tanmoy Chakraborty$^2$}\\
  $^1$Johns Hopkins University
  $^2$Indraprastha Institute of Information Technology - Delhi \\
  $^3$Sofia University
  $^4$Wipro AI Labs, India
  $^5$Qatar Computing Research Institute, HBKU, Doha \\
  \small\texttt{spraman3@jhu.edu}, \small\texttt{\{shivams, shad.akhtar, tanmoy\}@iiitd.ac.in}\\\small\texttt{mitko.bg.ss@gmail.com, pnakov@hbku.edu.qa}}
  
\date{}
 
\begin{document}
\setlength\abovedisplayskip{2pt}
\setlength{\belowdisplayskip}{2pt}
\maketitle
\begin{abstract}
Internet memes have become powerful means to transmit political, psychological, and socio-cultural ideas. Although memes are typically humorous, recent days have witnessed an escalation of \textit{harmful} memes used for trolling, cyberbullying, and abuse. Detecting such memes is challenging as they can be highly satirical and cryptic. Moreover, while previous work has focused on specific aspects of memes such as hate speech and propaganda, there has been little work on harm in general. Here, we aim to bridge this gap. We focus on two tasks: (\emph{i})~\textit{detecting harmful memes}, and (\emph{ii})~\textit{identifying the social entities they target}. We further extend a recently released \acldata\ dataset, which covered \emph{COVID-19}, with additional memes and a new topic: \emph{US politics}.
To solve these tasks, we propose \model\ (MultimOdal framework for detecting harmful MemEs aNd Their tArgets), a novel multimodal deep neural network that uses global and local perspectives to detect harmful memes. \model\ systematically analyzes the local and the global perspective of the input meme (in both modalities) and relates it to the background context. \model\ is interpretable and generalizable, and our experiments show that it outperforms several strong rivaling approaches.
\end{abstract}

\section{Introduction}

\begin{figure}[ht!]
\centering
\subfloat[{\centering Partially harmful meme.}\label{fig:meme:exm:1}]{
\includegraphics[width=0.23\textwidth, height=0.25\textwidth]{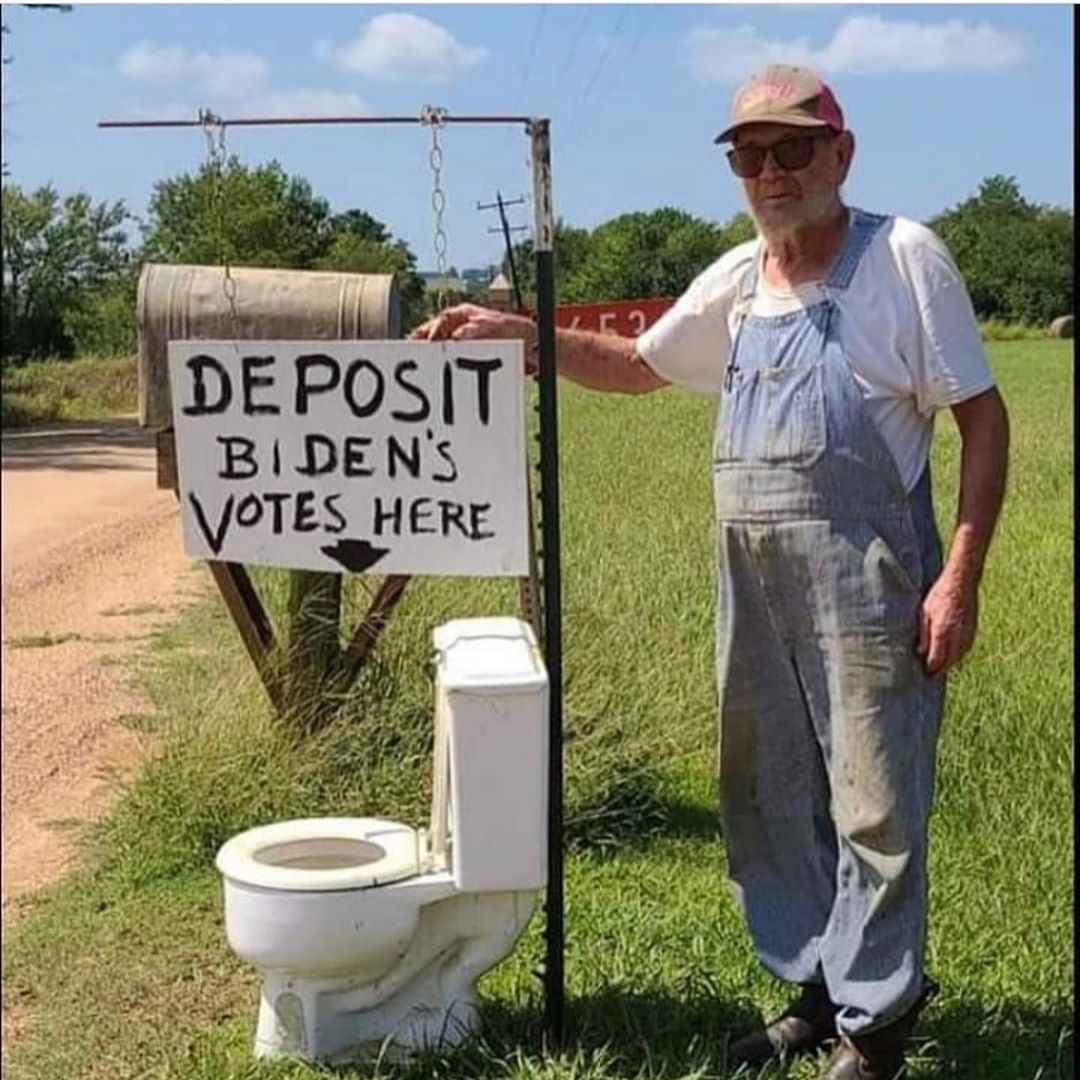}}\hspace{0.1em}
\subfloat[{Very harmful meme.}\label{fig:meme:exm:2}]{
\includegraphics[width=0.23\textwidth, height=0.25\textwidth]{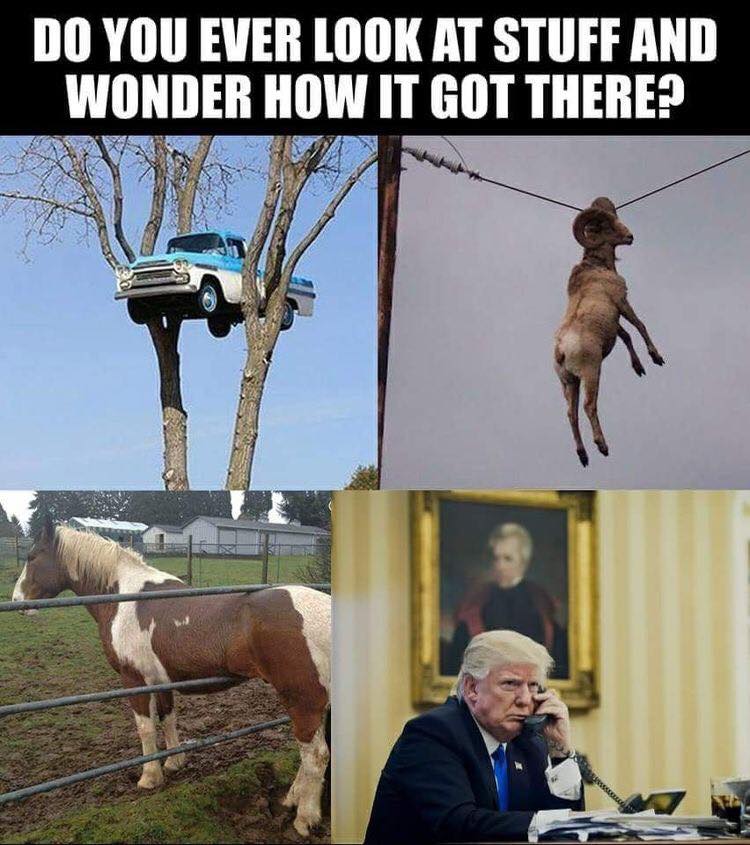}}

\caption{Examples of harmful memes.
(a) A meme that is \emph{partially harmful}, but is arguably not so hateful or offensive. (b) A meme, where the image and the text are not harmful when considered in isolation, but are \emph{very harmful} when taken as a whole.}
\label{fig:meme:example}
\end{figure}

The growing popularity of social media platforms has given rise to a new form of multimodal entity: the \textit{meme}, which is an image, embedded with a short piece of text. Memes are easily shared and can spread fast on the Internet, especially in social media. They are typically humorous and amusing in nature; however, by using an adroit combination of images and texts in the context of contemporary political and socio-cultural divisions, a seemingly harmless meme can easily become a multimodal source of harm.

Such harmful memes can be dangerous as they can easily damage the reputation of individuals, renowned celebrities, political entities, companies, or social groups, e.g.,~minorities. Despite memes being so influential, their multimodal nature and camouflaged semantics makes them very challenging to analyze.

The abundant quantity, fecundity and escalating diversity of online memes has led to a growing body of research on meme analysis, which has focused on tasks such as meme emotion analysis \cite{sharma-etal-2020-semeval, pramanick2021exercise}, sarcastic meme detection \cite{kumar2019sarc}, and hateful meme detection \cite{kiela2020hateful,zhou2021multimodal,velioglu2020detecting}. Research on these problems has shown that off-the-shelf multimodal systems, which often perform well on a range of visual-linguistic tasks, struggle when applied to memes. There are a number of reasons for that. First, memes are context-dependent, and thus focusing only on the image and on the text without background knowledge about the context in which the meme was generated, as well as some background information about people, companies and events, often is not enough to understand it.

Second, unlike other multimodal tasks, the image and the textual content in the meme are often uncorrelated, and its overall semantics is presented holistically. Finally, real-world memes can be noisy, and the text embedded in them can be hard to extract using standard OCR tools.  

The proliferation of virulent memes has stimulated research focusing on their dark sides: hate \cite{kiela2020hateful} and offensiveness \cite{suryawanshi-chakravarthi-2021-findings}. Recently, \citet{pramanick-etal-2021-detecting} defined the notion of \emph{harmful meme} and demonstrated its dependency on the background context. For example, the meme in Figure~\ref{fig:meme:exm:1} is somewhat harmful to Joe Biden in the context of an election, but it is arguably neither hateful nor offensive. Moreover, the notion of \textit{harm} is often apparent only when the two modalities are combined. For example, in Figure~\ref{fig:meme:exm:2}, the unimodal cues are not harmful, but the meme as a whole is harmful to Donald Trump. Moreover, identifying the target of harmful memes (e.g.,~\emph{Joe Biden} and \emph{Donald Trump}) requires separate analysis, which is not prevalent for hateful or offensive memes.    

With the above motivation in mind, here we aim to explore the role of background context for detecting harmful memes and for identifying the social entities they target. In particular, we make the following contributions: 

\begin{itemize}
    \item We extend our recently released \acldata\ dataset \cite{pramanick-etal-2021-detecting}, which covered \emph{COVID-19}, with aditional examples and a new topic (\emph{US Politics}), thus ending up with two datasets: \firstdata\ and \seconddata.
    \item We benchmark the two datasets against ten state-of-the-art unimodal and multimodal models, and we discuss the limitations of these models.
    \item We propose \model, a novel multimodal framework that systematically analyzes the local and the global perspective of the input meme and relates it to the background context, with the aim of detecting subtle harmful elements.
    \item 
   We perform extensive experiments on both datasets, and we show that \model\ outperforms the ten baselines in terms of accuracy by 1.3--2.6 points absolute for both tasks.
   \item Finally, we establish the generalizability and the interpretability of \model. 
\end{itemize}

\section{Related Work}

\subsection{Harm and Multimodality}

Various aspects of harm, such as hate speech, misinformation, and offensiveness, have been studied in isolation. \citet{Ahn2019MultimodalAF} addressed harmfulness in terms of obscenity and violence using multimodal approaches involving video and images. \citet{Hirsc}, \citet{9003892}, and \citet{INTERSPEECH2019:youtube} studied intentional deception and bias using textual and acoustic cues from the speech signal. \citet{8285382} and \citet{baly-etal-2020-written} designed robust systems for deception detection by combining acoustic, textual, and other information (visual, social). In recent work on detecting offensiveness in memes, \citet{suryawanshi-etal-2020-multimodal} showed improvements using an early-fusion multimodal approach that combines representations from unimodal models. Critical aspects such as prevalence of racial biases within the datasets and the modeling approaches were addressed in \cite{doi:10.1080/09500782.2018.1434787,bias_Davidson,Mozafari_2020,xia-etal-2020-demoting, zhou2021challenges}; they characterized the biases and proposed de-biasing mechanisms for tasks such as detecting toxic/abusive language and hate speech, as well as for identifying racial prejudices. 

Finally, recent research and a shared task focused on propaganda in memes \cite{ACL2021:propaganda:memes,SemEval2021:task6}, but did not target harmfulness per se.
 
\subsection{Harm and Memes}

There was a recent shared task on troll meme classification \cite{suryawanshi-chakravarthi-2021-findings}, and two tasks on hateful meme detection: \cite{kiela2020hateful} and \cite{zhou2021multimodal}. A number of models have been developed for these tasks. 
\citet{suryawanshi-chakravarthi-2021-findings} used a diverse set of models including logistic regression and BERT \cite{devlin2019bert}. \citet{muennighoff2020vilio} used a separate and a combined stream of Transformers \cite{vaswani2017attention}. \citet{velioglu2020detecting} used a Detectron-based representation to fine-tune Visual BERT \cite{li2019visualbert}, along with data augmentation. \citet{lippe2020multimodal} found UNITER \cite{chen2020uniter} to be a very strong choice for multimodal content. \citet{sandulescu2020detecting} used a multimodal deep ensemble, while examining both single-stream models such as ViLBERT \cite{lu2019vilbert}, VLP \cite{zhou2020unified}, and UNITER \cite{chen2020uniter}, and dual-stream models like LXMERT \cite{tan2019lxmert}. 

\citet{9253994} proposed a multimodal deep neural network with semantic and task-level attention for detecting medical misinformation. Another shared task, on memotion analysis \cite{sharma-etal-2020-semeval}, asked to recognize expressive emotions via sentiment (positive, negative, neutral), type of emotion (sarcastic, funny, offensive, motivation), and their intensity.
Recently, \citet{chandra2021subverting}
investigated antisemitism, its subtypes, and its use in memes. However, none of these studies addressed the broader concept of \emph{harmful memes}.

In our previous work \cite{pramanick-etal-2021-detecting}, we defined the notion of \textit{harmful meme}, and we differentiated it from hateful and offensive meme. We further formulated two tasks: (\emph{i})~detecting harmful meme, and (\emph{ii})~identifying the social entities they target. We also created \acldata, the first large-scale dataset for harmful meme analysis. However, \acldata\ contains memes related to only one topic, COVID-19. Here, we extend \acldata\ with additional examples and a new topic (\emph{US politics}). We further propose a novel multimodal framework, which systematically analyzes the local and the global perspective of the input meme (in both modalities) and relates it to the background context.

\subsection{Multimodal Pretraining}

Self-supervised pre-training using crossmodal and multimodal information saw an early reinstation with the work of \citet{NIPS2013_7cce53cf}, where semantic information from vast unannotated textual data was leveraged to classify images in a zero-shot setup. Similarly, Natural Language Processing (NLP) recently saw the emergence of Pattern-Exploiting Learning \cite{schick-schutze-2021-just}, which allows smaller models to outperform much larger ones such as \mbox{GPT-3}~\citep{brown2020language} when fine-tuned using a very small number of examples in a few-shot learning setup.

There have been also innovations towards better multimodal systems. \citet{ramesh2021zeroshot} proposed DALL-E, a simple yet scalable Transformer that autoregressively models the text tokens with the image features as a single stream of data, towards generating images from query texts and established competitive zero-shot performance. Then, \citet{radford2021learning} proposed a competitive model, CLIP, pre-trained on 400 million image--text pairs to train a joint multimodal visual-semantic embedding layer. In our experiments below, we compare our framework to CLIP and to variants thereof.

\section{Defining \emph{Harmful Meme}}
\label{sec:definition}

Following \citet{pramanick-etal-2021-detecting}, we abridge the definition of {\em harmful meme} as follows: a {\em multimodal unit consisting of an image and an embedded text that has the potential to cause harm to an individual, an organization, a community, or society}. Here, \emph{harm} includes mental abuse, defamation, psycho-physiological injury, socio-economic damages, proprietary damage, emotional disturbance, compensated public image, etc.

Offensive and hateful memes are harmful, but not the other way round. Offensive memes typically aim to mock or to bully a social entity, usually by using abusive words. A hateful meme contains derogatory content, influenced by utmost bias towards an entity (e.g.,~an individual a community, or an organization). The harmful content in a harmful meme is often camouflaged and might require critical judgment to detect. Furthermore, the social entities attacked or targeted by harmful memes can be any individual, organization, or community, as opposed to \emph{hateful} memes, where entities are attacked based on personal attributes.

\section{Data}
\label{sec:dataset}

Here, we describe our two datasets, \firstdata\ and \seconddata, which consist of memes related to COVID-19 and to US politics, respectively.

\subsection{Data Collection and Deduplication.}

To collect potentially harmful memes, we conducted keyword-based\footnote{\small \textbf{Example keywords for \firstdata:} \textit{Wuhan virus memes}, \textit{COVID vaccine memes}, \textit{Work from home memes}; \textbf{Example keywords for \seconddata:} \textit{Presidential debate memes}, \textit{Election-2020 vote counting memes}, \textit{Trump not wearing mask memes}.} web search on different sources, mainly Google Image. To alleviate potential biases from this search, we intentionally  included non-harmful examples using the same keywords. We used an extension\footnote{\url{download-all-images.mobilefirst.me}} of Google Chrome to download the images. We further scraped various publicly available meme pages on Reddit, Facebook, and Instagram. Unlike the Hateful Memes Challenge \cite{kiela2020hateful}, which offered synthetically generated memes, our datasets contain real-world memes. To remove noise, we maintained strict filtering on the resolution of the meme images and on the readability of the meme's text as part of the collection process. 
The process of filtering is described in detail in Appendix~\ref{sec:filtering}.

\begin{table*}[t]
\centering
    
\resizebox{\textwidth}{!}
{
\begin{tabular}{c l|r r r r |r r r r r}
\toprule
\multirow{2}{*}{\bf Dataset} & \multirow{2}{*}{\bf Split} & \multirow{2}{*}{\bf \#Memes} & \multicolumn{3}{c|}{\bf Harmfulness} &
\multirow{2}{*}{\bf \#Memes} &
\multicolumn{4}{c}{\bf Target}\\ \cline{4-6} \cline{8-11}
 
& & & \bf Very Harmful & \bf Partially Harmful & \bf Harmless & & \bf Individual & \bf Organization & \bf Community & \bf Society\\ \midrule
\multirow{4}{*}{\firstdata} & Train & 3,013 & 182 & 882 & 1,949 & 1,064 & 493 & 66 & 279 & 226\\
& Validation & 177 & 10 & 51 & 116 & 61 & 29 & 3 & 16 & 13 \\ 
& Test & 354 & 21 & 103 & 230 & 124 & 59 & 7 & 32 & 26 \\
\cline{3-11}
& \textbf{Total} & 3,544 & 213 & 1,036 & 2,295 & 1,249 & 582 & 75 & 327 & 265\\
\midrule
\multirow{4}{*}{\seconddata} & Train & 3,020 & 216 & 1,270 & 1,534 & 1,451 & 797 & 470 & 111 & 73\\
& Validation & 177 & 17 & 69 & 91 & 85 & 70 & 12 & 2 & 1\\ 
& Test & 355 & 25 & 148 & 182 & 170 & 96 & 54 & 12 & 8\\
\cline{3-11}
& \textbf{Total} & 3,552 & 258 & 1487 & 1,807 & 1,706 & 963 & 536 & 125 & 82 \\
\bottomrule
\end{tabular}}
\caption{Statistics about \firstdata\ and \seconddata. Harmful memes are also annotated with a target.}
\label{tab:dataset}
\end{table*}

To remove duplicate memes, we used two deduplication tools\footnote{\url{gitlab.com/opennota/findimagedupes}} \footnote{\url{github.com/arsenetar/dupeguru}} sequentially, and we preserved the memes with the highest resolution from each group of duplicates. The final size of \firstdata\ (on \emph{COVID-19}) and \seconddata (on \emph{US politics}) is 3,544 and 3,552 memes, respectively. We used the Google Cloud Vision API to extract the textual content of the memes.

\subsection{Data Annotation}

We followed the annotation procedure from \cite{pramanick-etal-2021-detecting}. In particular, we asked the annotators to label both the presence and the intensity of harm (\textit{harmful} vs. \textit{partially harmful}), as well as its target:

\begin{enumerate}  
   \item \textbf{Individual:} A person, usually a celebrity (e.g.,~a well-known politician, an actor, etc. such as \emph{Donald Trump}, \emph{Greta Thunberg}). 
   
    \item \textbf{Organization:} A group of people with a particular purpose, such as a business, a government department, a company, etc. Examples include research organizations such as \emph{WHO}, and political organizations such as the \emph{Democratic Party}.

    \item \textbf{Community:} A social unit with commonalities based on personal, professional, social, cultural, or political attributes such as religious views, country of origin, gender or gender identity, etc. 

    \item \textbf{Society:} When a meme promotes conspiracies or hate crimes, it is considered harmful to society. 
\end{enumerate}

We hired a total of 15 annotators: all of them experts in NLP or linguists, 22--40 years old, including 10 male and 5 female. We paid them fairly for their work as per the standard local pay rate.

Before the actual annotation process, we asked all annotators to go through the annotation guidelines. We further conducted several discussion sessions to evaluate whether they could understand what harmful content is and how to differentiate it from non-harmful content. The annotation process went through three stages: (\emph{i})~a dry run, (\emph{ii})~a final annotation, and (\emph{iii})~a consolidation stage.

\subsection{Inter-Annotator Agreement and Statistics}
The inter-annotator agreement (Cohen's $\kappa$) \cite{bobicev-sokolova-2017-inter} for harmfulness/target is 0.683/0.782 on \firstdata, and 0.675/0.790 on \seconddata, respectively. 
Table~\ref{tab:dataset} shows statistics about the data distribution and the split into train/validation/test sets, and Figure~~\ref{fig:source_label} (in the Appendix) shows statistics about the sources and the labels. Appendix~\ref{sec:annotation_guidelines} gives more detail about the target classes, the annotation guidelines, and the annotation process, and Appendix \ref{sec:lexical} offers some statistics about the textual content of the memes, including length distribution, and the most frequent words per dataset and per category.

\begin{figure*}[t!]
\centering
\hspace*{0cm}
  \includegraphics[width=0.85\linewidth]{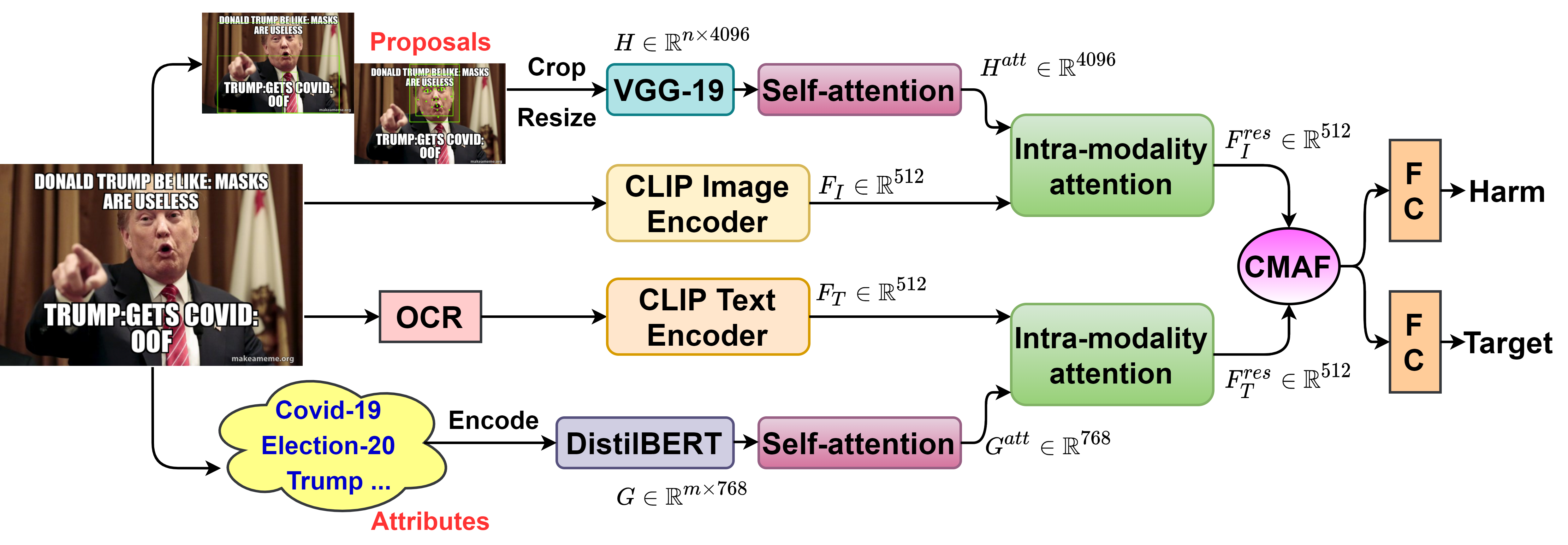}
  \caption{The architecture of our proposed model, \model.}
  \label{fig:model}
\end{figure*}

\section{\model: Our Proposed System}

Here, we describe our system, \model \ for harmful meme detection and target identification. It takes a meme as input, and extracts the embedded text using Google's OCR Vision API\footnote{\url{cloud.google.com/vision/docs/ocr}}. 
We encode each text--image pair using CLIP \cite{radford2021learning}, a pre-trained visual-linguistic model, leveraging its representations to capture the strong invariance and the overall semantics of the meme.

In addition to the CLIP features, we also identify faces and object proposals\footnote{Rectangular bounding boxes or regions of interests (ROI) surrounding the faces and the foreground objects.} \cite{NIPS2015_14bfa6bb}, and we extract various attributes (see Figure \ref{fig:attributes}), which define high-level topics or entities, such as \emph{Joe Biden} and the \emph{Republican Party}, in an image.

Then, on the visual side, we encode the ROIs using the pre-trained VGG-19 model \cite{simonyan2014very}, and on the textual side, we encode the topics/entities using DistilBERT \cite{sanh2019distilbert}. As we mentioned earlier, the analysis of harmful memes is challenging because of their abstruse nature without specific context. Thus, we hypothesize that adding object proposals and attributes would enable the model to understand the high-level concepts in the meme. Our analysis below shows that it indeed captures the appropriate background context reasonably well. Afterwards, we fuse the proposal and the attribute features together with the CLIP representations, separately for the image and for the textual representations, and we add first intramodel attentions, and then a crossmodal attention on top of them, in a hierarchical attention architecture. Finally, we use the resulting multimodal context-aware representation to predict the meme's harmfulness and its target. Figure~\ref{fig:model} shows the overall architecture of \model; below, we explain each of its components in detail.

\subsection{CLIP Representations}

CLIP (Contrastive Language–Image Pre-training) addresses the generalizability issues of standard computer vision systems \cite{simonyan2014very, he2016deep}, which are often good for some particular tasks, but perform poorly on stress sets and other tasks \cite{geirhos2018imagenet, alcorn2019strike, barbu2019objectnet}. CLIP is pre-trained using contrastive learning on 400M image--text pairs from the Internet. It offers excellent zero-shot capabilities due to the variety of images it has seen and the natural language supervision. In \model, given the meme's image $I$ and its OCR-extracted text $T$, we extract a CLIP image embedding $F_I$ and a CLIP text embedding $F_T$; both $F_I$ and $F_T$ are 512-dimensional vectors.

\subsection{Object Proposal and Attribute Representations}

\begin{figure}[!t]
\begin{subfigure}[t]{.45\textwidth}
\centering
\includegraphics[width=1\textwidth]{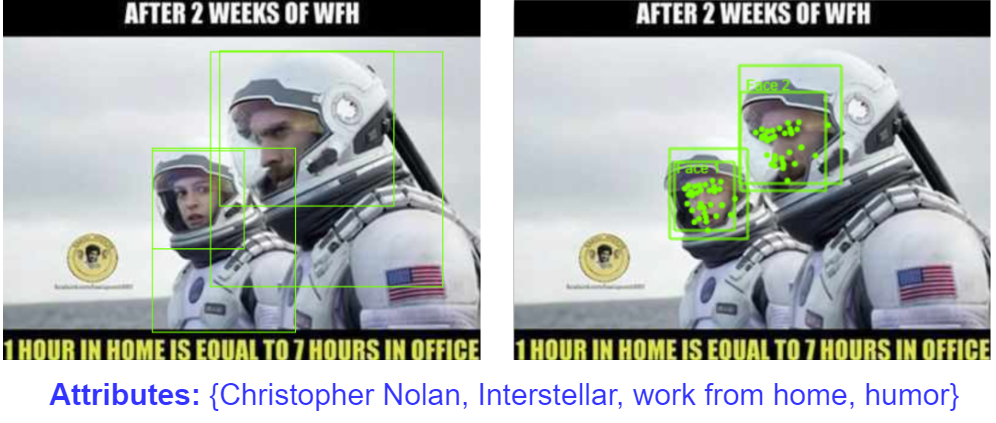}
\caption{Detected faces, foreground objects and image attributes for a \emph{harmless} meme from the \firstdata\ dataset.}
\label{fig:covid_attributes}
\end{subfigure}

\begin{subfigure}[b]{.45\textwidth}
\centering
\includegraphics[width=1\textwidth]{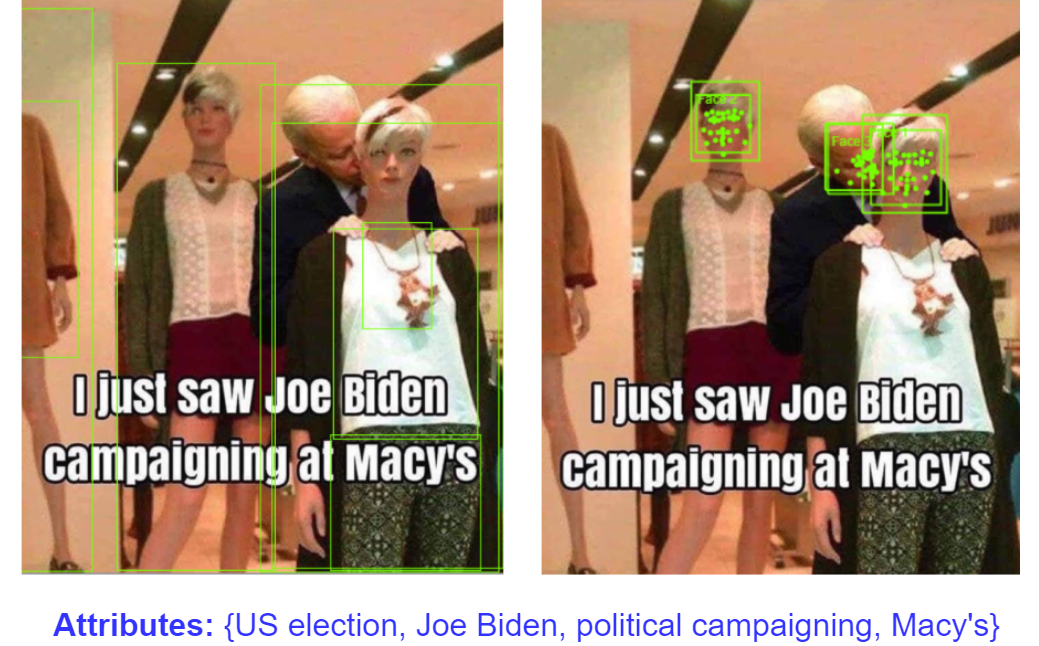}
\caption{Detected faces, foreground objects and image attributes for a \emph{very harmful} meme from the \seconddata\ dataset.}
\label{fig:pol_attributes}
\end{subfigure}
\caption{Detected proposals and attributes for two different memes from \firstdata\ and \seconddata\ datasets.}
\label{fig:attributes}
\end{figure}

Following previous studies \cite{wu2016value, cai2019multi} on image captioning and visual question answering, we introduce attributes as high-level image concepts in \model. Moreover, in addition to meme image attributes, we compute face and foreground object proposals, both of which help to capture subtle harmful contents and appropriate background context of the input meme. 
Figure~\ref{fig:attributes} shows the detected proposals and image attributes for two example memes. For the first meme (shown on Figure~\ref{fig:covid_attributes}), attributes such as \emph{Christopher Nolan} and \emph{Interstellar} capture the proper context, while for the second meme (shown on Figure~\ref{fig:pol_attributes}), the detected face of Joe Biden perceives minute harmful content.

We use three separate branches of the Google Cloud Vision API to detect faces,\footnote{\url{cloud.google.com/vision/docs/detecting-faces}} 
foreground objects,\footnote{\url{cloud.google.com/vision/docs/object-localizer}} and various image attributes.\footnote{\url{cloud.google.com/vision/docs/detecting-web}} 
Assume that given an input meme image $I$, the face and the object bounding boxes are $\{bb_1, bb_2, \dots, bb_n\}$, and the attributes are $\{att_1, att_2, \dots, att_m\}$. Each bounding box is cropped, reshaped and fed into VGG-19, which encodes it into a 4,096-dimensional representation. Next, we represent the encoded face and the detected object proposals as $H = \{h_1, h_2, \dots, h_n \}$, where $H \in \mathbb{R}^{n \times 4096}$. Similarly, each detected attribute is encoded and fed into DistilBERT to generate a 768-dimensional representation. We represent these attributes as $G = \{g_1, g_2, \dots, g_m \}$, where $G \in \mathbb{R}^{m \times 768}$. Note that the number of detected entities can vary.

\subsection{Intra-Modality Attention}

Next, we use self-attention over $n$ object proposals and $m$ image attributes to emphasize the most relevant ones for the target meme. The resulting self-attended respresentations are $H^{att}$ and $G^{att}$, respectively, where $H^{att} \in \mathbb{R}^{4096}$ and $H^{att} \in \mathbb{R}^{768}$.

\begin{eqnarray}
    H^{att}  =  \mathbf{W_H} \otimes H ; \hspace{3mm}
    G^{att}  =  \mathbf{W_G} \otimes G
\end{eqnarray}
where, $W_H \in \mathbb{R}^{1 \times n}$ and $W_G \in \mathbb{R}^{1 \times m}$ are learnable parameters, and $\otimes$ is a matrix outer product.

Subsequently, we fuse the self-attended object proposals with the CLIP image features in an intra-modality attention module. This stage aims to combine the local image descriptions with the global semantics of the meme. Similarly, we fuse the self-attended image attributes with the CLIP text features. Overall, the local and the global features capture the semantics of the meme considering the background context. Prior to the cross-modal attention fusion (CMAF) block, the proposal and the attribute features are projected to similar dimensions using a dense layer. 
\begin{eqnarray}
    F_I^{res} & = & \mathbf{W_I} \otimes [F_I, Dense(H^{att})]\\
    F_T^{att} & = & \mathbf{W_T} \otimes [F_T, Dense(G^{att})]
\end{eqnarray}

Finally we feed the resulting image and text features, $F_I^{res}, F_T^{att} \in \mathbb{R}^{512}$ into CMAF to obtain the final multimodal meme representation.

\subsection{Cross-Modality Attention Fusion}

For some memes, the text modality is more relevant, while for others, the image plays a crucial role. CMAF uses an attention mechanism to fuse the representations from the textual and the visual modalities. Motivated by \cite{gu2018hybrid}, we design our CMAF module with two major parts: modality attention generation and weighted feature concatenation. In the first part, we use a sequence of dense layers followed by a softmax layer to generate the attention scores $[a_v, a_t]$ for the two modalities. 

In the second part, we weigh the original unimodal features using their respective attention scores and we concatenate them together. We also use residual connections for better gradient flow.
\begin{eqnarray}
  F^{V}_{Meme} & = & (1 + a_v)F_I^{res} \\
  F^{T}_{Meme} & = & (1 + a_t)F_T^{res} \\
  F_{Meme}&=& \mathbf{W_F}\otimes [F^{V}_{Meme}, F^{T}_{Meme}]
\end{eqnarray}

\noindent where, $W_F \in \mathbb{R}^2$ is a learnable parameter, and $F_{Meme} \in \mathbb{R}^{512}$ is the final representation.

\subsection{Prediction and Training Objective}

We feed the final multimodal meme representation $F_{Meme}$ into two parallel fully-connected branches for the final classification: one branch per task. 

As there is class imbalance (shown in Table~\ref{tab:dataset} for each dataset and task), we use focal loss \cite{lin2017focal}, which down-weighs the easy examples and focuses training on the hard ones.

Finally, we train \model\ in a multi-task learning setup, where the loss for to target identification is considered only if the meme is \textit{partially harmful} or \textit{very harmful}. This is because we should not be looking for a target if the meme is not harmful.

\begin{table*}[hbtp]
\centering

\resizebox{\textwidth}{!}
{
\begin{tabular}{l | l | c c c | c c c | c c c | c c c}
\toprule
\multirow{3}{*}{\bf Modality} & \multirow{3}{*}{\bf Model} & \multicolumn{6}{c |}{\bf Harmful Meme Detection on \firstdata} & \multicolumn{6}{c}{\bf Harmful Meme Detection on \seconddata} \\

& &  \multicolumn{3}{c|}{\bf $2$-Class Classification} & \multicolumn{3}{c|}{\bf $3$-Class Classification} & \multicolumn{3}{c|}{\bf $2$-Class Classification} & \multicolumn{3}{c}{\bf $3$-Class Classification} \\

& & \bf Acc $\uparrow$ & \bf F1 $\uparrow$ & \bf MMAE $\downarrow$ & \bf Acc $\uparrow$ & \bf F1 $\uparrow$ & \bf MMAE $\downarrow$ & \bf Acc $\uparrow$ & \bf F1 $\uparrow$ & \bf MMAE $\downarrow$ & \bf Acc $\uparrow$ & \bf F1 $\uparrow$ & \bf MMAE $\downarrow$ \\

\midrule

& Human$^\dagger$ & 90.68 & 83.55 & 0.1723 & 86.10 & 65.10 & 0.4857 & 94.40 & 88.47 & 0.1028 & 92.12 & 70.35 & 0.6274 \\ 

& Majority & 64.76 & 39.30 & 0.5000 &64.76 & 26.20 & 1.0000 & 51.27 & 33.39 & 0.5000 & 51.27 & 22.59 & 1.0000 \\
\hline

\multirow{1}{3cm}{{\tt Text (T)} Only} & TextBERT & 70.17 & 66.25 & 0.2911 & 68.93 & 48.72 & 0.5591 & 80.12 & 78.35 & 0.1660 & 74.55 & 54.08 & 0.7742 \\

\hline

\multirow{4}{3cm}{{\tt Image (I)} Only} & VGG19 & 68.12 & 61.86 & 0.3190 & 66.24 & 41.76 & 0.6487 & 70.65 & 70.46 & 0.1887 & 73.65 & 51.89 & 0.8466 \\
& DenseNet-161 & 68.42 & 62.54 & 0.3125 & 65.21 & 42.15 & 0.6326 & 74.05 & 73.68 & 0.1845 & 71.80 & 50.98 & 0.8388 \\
& ResNet-152 & 68.74 & 62.97 & 0.3114 & 65.29 & 43.02 & 0.6264 & 73.14 & 72.77 & 0.1800 & 71.02 & 50.64 & 0.8900\\
& ResNeXt-101 & 69.79 & 63.68 & 0.3029 & 66.55 & 43.68 & 0.6499 & 73.91 & 73.57 & 0.1812 & 71.84 & 51.45 & 0.8422 \\

\hline

\multirow{3}{3cm}{{\tt I} + {\tt T} (Unimodal Pre-training)} & Late Fusion & 73.24 & 70.25 & 0.2927 & 66.67 & 45.06 & 0.6077 & 78.26 & 78.50 & 0.1674 & 76.20 & 55.84 & 0.7245\\
& Concat BERT & 71.82 & 71.82 & 0.3156  & 65.54 & 43.37 & 0.5976 & 77.25 & 76.38 & 0.1743 & 76.04 & 55.95 & 0.7450 \\
& MMBT & 73.48 & 67.12 & 0.3258 & 68.08 & 50.88 & 0.6474 & 82.54 & 80.23 & 0.1413 & 78.14 & 58.03 & 0.7008\\

\hline

\multirow{2}{3cm}{{\tt I} + {\tt T} (Multimodal Pre-training)}
& ViLBERT CC & 78.53 & 78.06 & 0.1881 & \bf 75.71 & 48.82 & 0.5329 & \bf 87.25 & \bf 86.03 & \bf 0.1276 & \bf 84.66 & \bf 64.70 & \bf 0.6982  \\
& V-BERT COCO & \bf 81.36 & \bf 80.13 & \bf 0.1857 & 74.01 & \bf 53.85 & \bf 0.5303 & 86.80 & 86.07 & 0.1318 & 84.02 & 63.68 & 0.7020\\

\midrule

\multirow{5}{3cm}{Proposed System and Variants}
& CLIP & 74.23 & 73.85 & 0.2955 & 67.04 & 44.25 & 0.6228 & 80.55 & 80.25 & 0.1659 & 77.00 & 56.85 & 0.7852 \\
& CLIP + Proposals & 77.65 & 76.90 & 0.2142 & 70.52 & 45.60 & 0.5955 & 84.16 & 83.80 & 0.1556 & 81.06 & 60.65 & 0.7122\\
& CLIP + Attributes & 78.10 & 77.64 & 0.2010 & 71.05 & 45.55 & 0.5887 & 84.02 & 83.85 & 0.1508 & 80.75 & 60.23 & 0.7058 \\
& \model\ w/o {\tt CMAF} & 80.75 & 80.17 & 0.1896 & 74.85 & 51.25 & 0.5360 & 86.20 & 85.55 & 0.1355 & 83.85 & 63.02 & 0.6990\\
& \model\ & \bf 83.82 & \bf 82.80 & \bf 0.1743 & \bf 77.10 & \bf 54.74 & \bf  0.5132 & \bf 89.84 & \bf 88.26 & {\bf 0.1314} & \bf 87.14 & \bf 66.66 & \bf 0.6805 \\

\hline

\multicolumn{2}{c|}{\bf {$\Delta_{\model-best\_model}$}} & \textcolor{blue}{2.46} & \textcolor{blue}{2.67} & \textcolor{blue}{0.0114} & \textcolor{blue}{1.39} & \textcolor{blue}{0.89} & \textcolor{blue}{0.0171} & \textcolor{blue}{2.59} & \textcolor{blue}{2.23} & \textcolor{red}{0.0038} & \textcolor{blue}{2.48} & \textcolor{blue}{1.96} & \textcolor{blue}{0.0177}\\

\bottomrule
\end{tabular}}
\caption{Performance on the two tasks. For two-class, we merge \textit{very harmful} and \textit{partially harmful}. $^\dagger$This row shows the human performance on test, and the last row shows the improvement of \model\ over the best baseline.}
\label{tab:results_harmful}
\end{table*}

\section{Experiments}

We train \model\ and all baselines using Pytorch on NVIDIA Tesla V100 GPU, with 32 GB dedicated memory, with CUDA-11.2 and cuDNN-8.1.1 installed. The hyper-parameter values for all models are given in Appendix~\ref{hyperparameters}.

We experiment with \firstdata\ and \seconddata\ using a variety of state-of-the-art unimodal textual models, unimodal visual models, and multimodal models that were pre-trained on both modalities. We use three measures for evaluation: Accuracy, Macro-F1, and Macro-Averaged Mean Absolute Error (MMAE) \citep{baccianella2009evaluation}. For the first two, higher values are better, while for MMAE, lower values are better. 

\subsection{Baselines}

\subsubsection{Unimodal Models}
$\rhd$ \textbf{Text BERT:} We use BERT \cite{devlin2019bert} as our unimodal text-only model. \\
$\rhd$ \textbf{VGG19, DenseNet, ResNet, ResNeXt:} For the unimodal visual-only models, we use four well-known models: VGG19 \cite{simonyan2014very}, DenseNet-161 \citep{huang2017densely}, ResNet-152 \citep{he2016deep}, and ResNeXt-101 \citep{xie2017aggregated}, pre-trained on ImageNet \citep{deng2009imagenet}.

\subsubsection{Multimodal Models}
    $\rhd$ \textbf{Late fusion:} This model uses the average prediction scores of ResNet-152 and BERT.
\\    
    $\rhd$ \textbf{Concat BERT:} This model concatenates the representations from ResNet-152 and BERT, and uses a perceptron as a classifier on top of them.
\\    
    $\rhd$ \textbf{MMBT:} This is a Multimodal Bitransformer \cite{kiela2020supervised}, capturing the intra-modal and the inter-modal dynamics of the two modalities. 
\\    
    $\rhd$ \textbf{ViLBERT CC:} Vision and Language BERT \cite{lu2019vilbert}, trained on an intermediate multimodal objective (conceptual captions) \cite{sharma2018conceptual}, is a strong model with task-agnostic joint representation of image and text.   
\\    
    $\rhd$ \textbf{Visual BERT COCO:} This is Visual BERT \citep{li2019visualbert}, pre-trained on the COCO dataset \citep{lin2014microsoft}, another strong multimodal model.

\section{Experimental Results}

We compare \model\ to unimodal textual models, unimodal visual models, and multimodal models pre-trained on both modalities.

Except for the unimodal visual models, we use the MMF framework.\footnote{\url{github.com/facebookresearch/mmf}} We further explore the generalizability and the interpretability of \model.

\subsection{Harmful Meme Detection}

Table \ref{tab:results_harmful} shows the results for harmful meme detection. We start by merging the \textit{partially harmful} and the \textit{very harmful} classes, thus ending up with binary classification. In both datasets, \textit{harmless} is the majority class; the majority class baseline yields accuracy of 64.76 on \firstdata\ and of 51.27 on \seconddata. Among the unimodal models, those using the textual modality perform better. In case of \firstdata, the accuracy of the unimodal models is 68.1--70.2, while on \seconddata, it is 70.7--80.1. 

We also see that multimodal models outperform unimodal ones, and more sophisticated fusion techniques perform better. For example, late fusion, the simplest one, performs only slightly better than unimodal models, while MMBT, yields 2.5--3.3 absolute points of improvement. We also notice the effectiveness of multimodal pre-training. On \mbox{\firstdata}, Visual BERT COCO outperforms all other models, while on \seconddata, VilBERT CC is the best. Thus, below we will compare the performance of \model\ to these two models.

In the binary case, \model\ achieves 2.46 absolute points of improvement on \firstdata, and 2.59 points on \seconddata\ over the best models. The corresponding Macro-F1 scores also improve by a similar margin. We show in Section~\ref{sec:ablation} that all modules in \model\ contribute to this.

For the 3-class harmful meme detection, we see a similar trend: early-fusion models with multimodal pre-training (ViLBERT CC, V-BERT COCO) outperform unimodal and simple multimodal ones. Moreover, \model\ achieves an improvement of 1.39 and 2.48 points absolute over the corresponding best models on \firstdata\ and \seconddata.  

\begin{table}[t]
\centering
\resizebox{1\columnwidth}{!}
{
\begin{tabular}{l | l | c c c | c c c}
\toprule
\multirow{2}{*}{\bf Modality} & \multirow{2}{*}{\bf Model} & \multicolumn{3}{c|}{\bf Target on \firstdata} & \multicolumn{3}{c}{\bf Target on \seconddata}  \\

& & \bf Acc $\uparrow$ & \bf F1 $\uparrow$ & \bf MMAE $\downarrow$ & \bf Acc $\uparrow$ & \bf F1 $\uparrow$ & \bf MMAE $\downarrow$ \\

\midrule

& Human$^\dagger$ & 87.55 & 82.01 & 0.3647 & 90.58 & 72.68 & 0.6324\\
& Majority & 46.60 & 15.89 & 1.5000 & 56.47 & 18.05 & 1.5000\\

\hline

\multirow{1}{3cm}{Text (T) only} & TextBERT & 69.35 & 55.60 & 0.8988 & 72.54 & 60.36 & 0.8895\\

\hline

\multirow{4}{3cm}{Image (I) only} & VGG19 & 63.48 & 53.60 & 1.0549 & 68.24 & 55.24 & 1.0225\\
& DenseNet-161 & 64.52 & 53.51 & 1.0065 & 69.40 & 57.95 & 0.9540\\
& ResNet-152 & 65.75 & 53.78 & 1.0459 & 68.75 & 57.00 & 0.9667\\
& ResNeXt-101 & 65.82 & 53.95 & 0.9277 & 70.22 & 59.67 & 0.9245\\

\hline

\multirow{3}{3cm}{I + T (Unimodal Pretraining)} & Late Fusion & 72.58 & 58.43 & 0.6318 & 73.25 & 64.28 & 0.8541\\ 
& Concat BERT & 67.74 & 49.77 & 0.8879 & 72.46 & 60.87 & 0.8655\\
& MMBT & 72.58 & 58.35 & 0.6318 & 74.65 & 65.12 & 0.8441\\

\hline

\multirow{2}{3cm}{I + T (Multimodal Pretraining)}
& ViLBERT CC & 72.58 & 57.17 & 0.8035 & 77.25 & \bf 67.39 & \bf 0.8410\\
& V-BERT COCO & \bf 75.81  & \bf 65.77 & \bf 0.5036 & \bf 77.28 & 66.90 & 0.8536\\

\midrule

\multirow{5}{3cm}{Proposed System and Variants}
& CLIP & 72.47 & 62.14 & 0.6312 & 72.40 & 65.66 & 0.8557\\
& CLIP + Proposals & 74.85 & 64.38 & 0.5746 & 75.85 & 66.13 & 0.8482 \\
& CLIP + Attributes & 74.56 & 61.38 & 0.6015 & 76.20 & 66.34 & 0.8491\\
& \model\ w/o {\tt CMAF} & 76.16 & 64.80 & 0.5422 & 77.54 & 67.25 & 0.8430 \\
& \model\ & \bf 77.95 & \bf 69.65 & \bf 0.4225 & \bf 78.54 & \bf 68.83 & \bf 0.8295 \\

\hline 

\multicolumn{2}{c|}{\bf {$\Delta_{\model-best\_model}$}} & \textcolor{blue}{2.14} & \textcolor{blue}{3.88} & \textcolor{blue}{0.0811} & \textcolor{blue}{1.26} & \textcolor{blue}{1.44} & \textcolor{blue}{0.0115} \\

\bottomrule
\end{tabular}}
\caption{Performance for target identification. $^\dagger$This row shows the human performance on the test set.
}
\label{tab:results_target}
\end{table}

\subsection{Target Identification}

Table \ref{tab:results_target} shows the performance for the 4-class target identification. Here, \seconddata\ is more imbalanced than \firstdata. The majority class baseline yields accuracy of 46.60 on \firstdata\ and of 56.47 on \seconddata. Similarly to the earlier task, unimodal models perform poorly, achieving 63.4--69.3 accuracy on \firstdata, and 68.2--72.5 on \mbox{\seconddata}. Adding multimodal cues with multimodal pre-training yields sizable improvements. Visual BERT COCO is the best on \firstdata, and ViLBERT CC is the best on \seconddata. However, \model\ outperforms the best models by 2.14 points absolute in terms of accuracy and by 3.88 points in terms of F1 score on \firstdata, and by 1.26 points of accuracy and 1.44 points of F1 on \seconddata.

\begin{table*}[t]
\centering

\resizebox{0.75\textwidth}{!}{
\begin{tabular}{c l| ccc|ccc|ccc}
\toprule
& & \multicolumn{3}{c|}{\bf \firstdata} & \multicolumn{3}{c|}{\bf \seconddata} & \multicolumn{3}{c}{\bf Combined} \\
& & \bf H-2$\dagger$ & \bf H-3$\ddagger$ & \bf  Tar$^\star$ & \bf H-2$\dagger$ & \bf H-3$\ddagger$ & \bf Tar$^\star$ & \bf  H-2$\dagger$ & \bf H-3$\ddagger$ & \bf Tar$^\star$ \\
\midrule

\multirow{3}{*}{\bf \firstdata} & ViLBERT & 78.06 & 48.82 & 57.17 & 74.20 & 51.39 & 54.10 & 74.85 & 44.15 & 46.52 \\
& V-BERT & 80.13 & 53.85 & 68.77 & 74.56 & 52.87 & 53.46 & 75.04 & 45.20 & 47.66 \\
& \model & \bf 82.80 & \bf 54.74 & \bf 69.65 & 80.25 & 61.87 & 58.39 & \textcolor{blue}{81.66} & \textcolor{blue}{49.83} & 50.12 \\
\hline 
\multirow{3}{*}{\bf \seconddata} & ViLBERT & 71.28 & 42.57 & 48.20 & 86.03 & 64.70 & 67.39 & 75.88 & 44.18 & 45.82 \\
& V-BERT & 72.58 & 45.10 & 54.07 & 86.07 & 63.68 & 66.90 & 76.20 & 45.69 & 47.38 \\
& \model & 76.30 & 50.46 & 58.33 &\bf 88.26 & \bf 66.66 & \bf 68.83 & 80.75 & 49.70 & \textcolor{blue}{50.28}\\
\hline 
\multirow{3}{*}{\bf Combined} & ViLBERT & 73.48 & 43.11 & 51.45 & 76.92 & 56.50 & 60.20 & 79.20 & 53.65 & 58.12 \\
& V-BERT & 74.88 & 46.28 & 60.82 & 76.85 & 56.07 & 58.22 & 80.45 & 53.98 & 58.76 \\
& \model & \textcolor{blue}{79.50} & \textcolor{blue}{51.07} & \textcolor{blue}{62.56} & \textcolor{blue}{81.09} & \textcolor{blue}{62.85} & \textcolor{blue}{61.87} & \bf 85.20 & \bf 58.44 & \bf 61.20 \\
\bottomrule

\end{tabular}}
\caption{Transferability of the two best-performing baselines and \model\ on \firstdata, on \seconddata, and on the combination thereof. The models are trained on the dataset in the row and tested on the one in the column. All scores are Macro F1. H-$2\dagger$ is $2$-class, and H-$3\ddagger$ is $3$-class harmful meme detection, and Tar$^\star$ is target identification. The blue color indicates the best transferable results for each task--dataset combination.}
\label{tab:transferability}
\end{table*}

\subsection{Human Evaluation}

In order to understand how humans perceive these tasks compared to neural systems, we performed an extensive manual evaluation. We hired three linguists to label the test set. The results show that for all three experiments and on both datasets, \model \ is the best performing model; yet, it is 4.5--12 absolute points of accuracy behind human performance. The difference is more sizable for target identification, indicating the difficulty of that task and suggesting that there is a lot of room for further improvement.  

\subsection{Ablation Study} \label{sec:ablation}

Here, we present an ablation study, analyzing the contribution of each of the components of \model: proposal, attribute detection blocks, and CMAF module. The last five rows of Tables \ref{tab:results_harmful} and \ref{tab:results_target} show these results. 

We can see in the tables that without the proposals and the attributes, CLIP performs similarly to MMBT. Then, adding the proposals improves accuracy by 2.3--4.1 points absolute for all tasks and datasets. Incorporating attributes in CLIP also improves results by a similar margin. When both are added, the accuracy improves further, outperforming CLIP by 5.5--10 points absolute in terms of accuracy and by 3.2--10.5 points absolute in terms of F1 score. The CMAF module also plays a pivotal role: we notice a drop of 1--3.8 points absolute in terms of accuracy, and 1.5--4.8 points absolute in terms of F1 score when CMAF is replaced by simple concatenation. Hence, we can conclude that each block of \model's architecture helps to boost the overall performance.

\begin{figure}[t]
\centering
\subfloat[{LIME image - \model.}\label{fig:lime_image_GLARE}]{
\includegraphics[width=0.22\textwidth, height=0.22\textwidth]{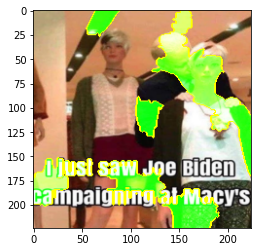}}
\hspace{0.5em}
\subfloat[{LIME image - ViLBERT}\label{fig:lime_image_ViLBERT}]{
\includegraphics[width=0.22\textwidth, height=0.22\textwidth]{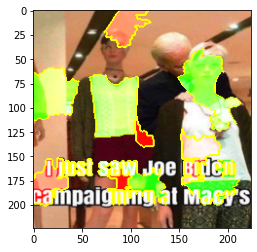}}
\\
\vspace{0.5em}
\subfloat[{LIME text - \model.}\label{fig:lime_text_GLARE}]{
\includegraphics[width=\columnwidth]{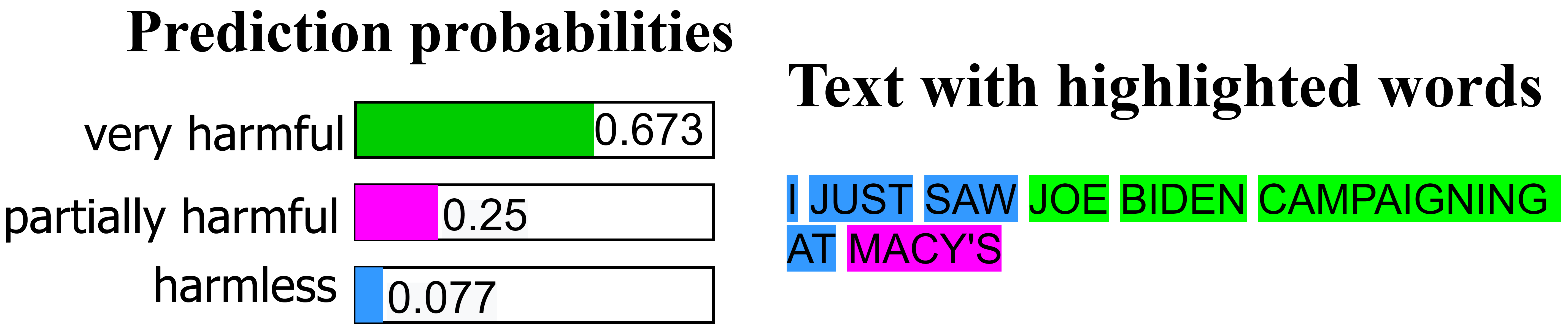}}

\caption{Visualization of explanation as generated by LIME on both modalities for \model\ and ViLBERT.}
\label{fig:lime_explainability}
\end{figure}

\begin{figure}[t]
\centering
\subfloat[{Misclassified meme.}\label{fig:misclassified_meme}]{
\includegraphics[width=0.22\textwidth, height=0.22\textwidth]{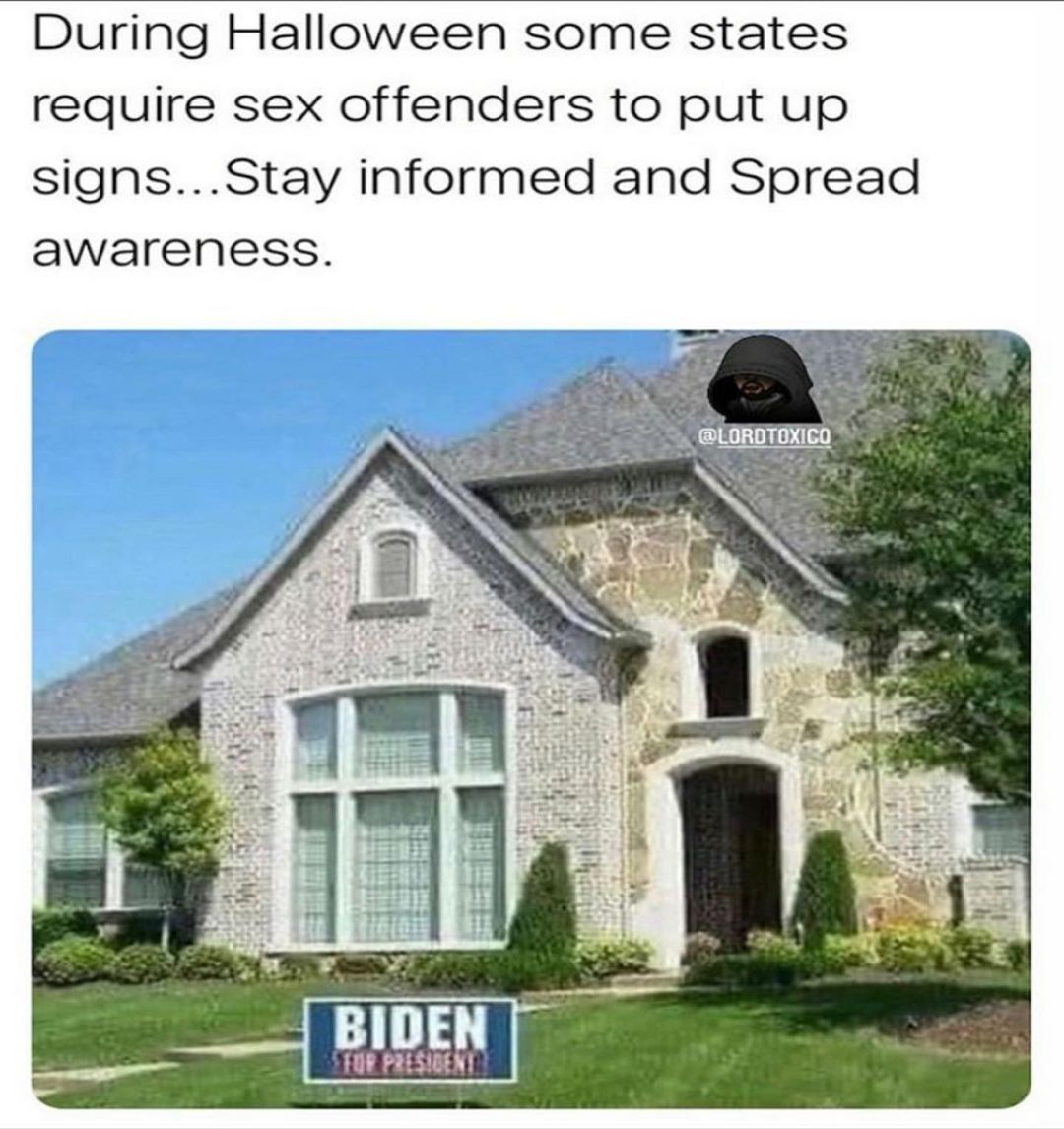}}
\hspace{0.5em}
\subfloat[{LIME image - \model.}\label{fig:misclassified_lime}]{
\includegraphics[width=0.22\textwidth, height=0.22\textwidth]{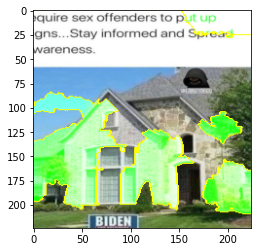}}

\caption{Misclassified example and explanation.}
\label{fig:lime_error}
\end{figure}

\subsection{Transferability of \model}

Table \ref{tab:transferability} shows the transferability of \model\ on \firstdata, on \seconddata, and on the combination thereof, compared to ViLBERT and Visual BERT.
We can see that when training and testing on the same dataset, all models yield high F1 scores. However, when trained on one dataset and tested on a different one, \model\ yields 2.2--6.6, 1.1--9.0, and 0.9--4.2 points of absolute improvements in terms of F1 score for 2-class and 3-class harmful meme detection and for target identification, respectively. CLIP, which was pre-trained on 400M image--text pairs, contributes to the superior transferability of \model.

\subsection{Side-by-Side Diagnostics}

We visualize the explainability of \model\ and we compare it to ViLBERT using LIME \citep{ribeiro2016should}. We take the example from Figure~\ref{fig:pol_attributes} for our analysis. \model\ correctly classified it as \textit{very harmful} with a dominant probability of 0.673, but ViLBERT fails. Figures~\ref{fig:lime_image_GLARE} and \ref{fig:lime_image_ViLBERT} highlight the most important super-pixels contributing to the decision of \model\ and ViLBERT, respectively. We notice that the face of Joe Biden and the mannequin, which are presented in a very insulting way in this meme, contribute heavily to the prediction of \model. However, as Biden's face is partially occluded, ViLBERT cannot recognize this harmful gesture. The fine-grained face detection and the robust CLIP encoder help \model\ to identify this subtle harmful element. Figure~\ref{fig:lime_text_GLARE} shows the contribution of different words in the meme's text to the prediction of \model. Overall, the word \emph{CAMPAIGNING} and the conflicting gesture in the image make the meme \textit{very harmful}.

\subsection{Error Analysis}

Figure \ref{fig:misclassified_meme} shows a \textit{very harmful} meme on which \model\ fails: the image contains no harmful gestures, and the text has no harmful words. The most contributing super pixels are also spread randomly, as shown in Figure~\ref{fig:misclassified_lime}. Moreover, the detected attributes, \{\textit{palace}, \textit{summer}, \textit{mansion}\}, do not model context well. Yet, the entrenched semantics of the entire meme makes it \textit{very harmful}.

\section{Conclusion and Future Work}

We introduced two large-scale datasets, \firstdata\ and \seconddata, for detecting harmful memes and their targets. We further proposed \model, a novel multimodal deep neural network that systematically analyzes the local and the global perspective of the input meme (in both modalities) and relates it to the background context. Extensive experiments on the two datasets showed the efficacy of \model, which outperforms ten baselines for both tasks. We further demonstrated its transferability and interpretability.

In future work, we plan to extend the datasets with more domains and languages.

\section*{Acknowledgments}
The work was partially supported by a Wipro research grant, the Infosys Centre for AI, IIIT Delhi, India, and ihub-Anubhuti-iiitd Foundation, set up under the NM-ICPS scheme of the Department of Science and Technology, India. 

It is also part of the Tanbih mega-project, developed at the Qatar Computing Research Institute, HBKU, which aims to limit the impact of ``fake news,'' propaganda, and media bias by making users aware of what they are reading.

\section*{Ethics and Broader Impact}

\paragraph{Reproducibility.}  We present detailed hyper-parameter configurations in Table \ref{tab:hyperparameters} and Appendix \ref{hyperparameters}. 
The source code, and the datasets (\firstdata and \seconddata) are available at \url{http://github.com/LCS2-IIITD/MOMENTA}
 
\paragraph{User Privacy.}

Our datasets only include memes, and they do not contain any user information. All the memes in our datasets were collected from publicly available web pages and there are no known copyright issues regarding them. The sources are listed in Section~\ref{sec:dataset} and Figure~\ref{fig:source_label_statistics}. Note that we also release links to the memes instead of the actual memes. In this way, we ensure that if a user deletes a posted meme, that meme would not be available in our datasets anymore. The same strategy was previously used by several researchers to distribute collections of tweets.

\paragraph{Annotation.}

The annotation was conducted by the experts working in NLP or linguists in India. We treated the annotators fairly and with respect. They were paid as per the standard local paying rate. Before beginning the annotation process, we requested every annotator to thoroughly go through the annotation guidelines. We further conducted several discussion sessions to make sure all annotators could understand well what harmful content is and how to differentiate it from humorous, satirical, hateful, and non-harmful content.

\paragraph{Biases.}

Any biases found in the dataset are unintentional, and we do not intend to cause harm to any group or individual. We note that determining whether a meme is harmful can be subjective, and thus it is inevitable that there would be biases in our gold-labeled data or in the label distribution. We address these concerns by collecting examples using general keywords about COVID-19, and also by following a well-defined schema, which sets explicit definitions during annotation. Our high inter-annotator agreement makes us confident that the labeling of the data is correct most of the time.

\paragraph{Misuse Potential.}

We ask researchers to be aware that our dataset can be maliciously used to unfairly moderate memes based on biases that may or may not be related to demographics and other information within the text. Intervention with human moderation would be required in order to ensure that this does not occur.

\paragraph{Intended Use.}

We release our dataset aiming to encourage research in studying harmful memes on the web. We distribute the dataset for research purposes only, without a license for commercial use. We believe that it represents a useful resource when used in the appropriate manner.

\paragraph{Environmental Impact.}

Finally, we would also like to warn that the use of large-scale Transformers requires a lot of computations and the use of GPUs/TPUs for training, which contributes to global warming \cite{strubell2019energy}. This is a bit less of an issue in our case, as we do not train such models from scratch; rather, we fine-tune them on relatively small datasets. Moreover, running on a CPU for inference, once the model has been fine-tuned, is perfectly feasible, and CPUs contribute much less to global warming.

\bibliography{custom}
\bibliographystyle{acl_natbib}

\newpage
\thispagestyle{plain}
\makeatletter
\twocolumn[\LARGE \bf \centering Appendix \par \bigskip
 ]
\appendix

\counterwithin{figure}{section}
\numberwithin{table}{section}

\section{Implementation Details and Hyperparameter Values} \label{hyperparameters}

We train all the models using Pytorch on an NVIDIA Tesla V100 GPU, with 32 GB dedicated memory, CUDA-11.2 and cuDNN-8.1.1 installed. For the unimodal models, we import all the pre-trained weights from the TORCHVISION.MODELS\footnote{\url{http://pytorch.org/docs/stable/torchvision/models.html}} subpackage of the PyTorch framework. We initialize the remaining weights randomly using a zero-mean Gaussian distribution with a standard deviation of 0.02. Statistics about the dataset are shown in Table~\ref{tab:dataset}, where we can see that there is label imbalance both for harmfulness intensity ([\textit{Very Harmful}, \textit{Partially Harmful}] vs. \textit{Harmless}) and for target classification ([\textit{Individual}, \textit{Organization}, \textit{Community}] vs. \textit{Entire Society}). We address this using focal loss (FL) \cite{lin2017focal}, which down-weighs the easy examples and focuses training on the hard ones. We train \model\ in a multi-task learning setup, where the loss due to target identification is considered only if the meme is \textit{partially harmful} or \textit{very harmful}. 

We train all models we experiment with using the Adam optimizer \cite{kingma2014adam} and a negative log-likelihood loss (NLL) as the objective function. Table~\ref{tab:hyperparameters} gives more detail about the hyper-parameters we used for training.

\begin{table*}[h]
\centering
\resizebox{0.78\textwidth}{!}
{
\begin{tabular}{c  l  c  c  c  c  c  c }
\toprule
& &  \bf Batch-size & \bf Epochs & \bf Learning Rate & \bf Image Encoder & \bf Text Encoder & \bf \#Param \\
\midrule

\multirow{5}{*}{\begin{turn}{-90} \centering \bf Unimodal \end{turn}} & TextBERT & 16 & 20 & 0.001 & - & Bert-base-uncased & 110M\\
& VGG19 & 64 & 200 & 0.01 & VGG19 & - & 138M\\
& DenseNet-161 & 32 & 200 & 0.01 & DenseNet-161 & - & 28M\\
& ResNet-152 & 32 & 300 & 0.01 & ResNet-152 & - & 60M\\
& ResNeXt-101 & 32 & 300 & 0.01 & ResNeXt-101 & - & 83M\\
\midrule
\multirow{6}{*}{\begin{turn}{-90} \centering \bf Multimodal \end{turn}} & Late Fusion & 16 & 20 & 0.0001 & ResNet-152 & Bert-base-uncased & 170M \\
& Concat BERT & 16 & 20 & 0.001 & ResNet-152 & Bert-base-uncased & 170M\\
& MMBT & 16 & 20 & 0.001 & ResNet-152 & Bert-base-uncased & 169M\\
& ViLBERT CC & 16 & 10 & 0.001 & Faster RCNN & Bert-base-uncased & 112M\\
& V-BERT COCO & 16 & 10 & 0.001 & Faster RCNN & Bert-base-uncased & 247M\\ \cline{2-8}
& \model & 64 & 50 & 0.001 & VGG19 & DistilBERT-base-uncased & 358M \\

\bottomrule
\end{tabular}}

\caption{Values of the hyperparameters for the models we experimented with.}

\label{tab:hyperparameters}
\end{table*}

\section{Data Filtering}\label{sec:filtering}

We apply the following fine-grained filtering criteria during data collection and annotation for the examples we include in our datasets, \firstdata\ and \seconddata:

\begin{enumerate}
    \item The meme text must be in English; no other languages and no code-switching are allowed.
    \item The meme text must be readable. Thus, we exclude blurry text, incomplete text, etc.
    \item The meme must contain no cartoons (as they are often very hard to interpret by AI systems).
    \item The meme must be multimodal, i.e.,~it should contain both an image and soem text.
\end{enumerate}

\begin{figure*}[ht!]
\centering
\subfloat[\label{fig:meme_rej:exm:1}]{
\includegraphics[width=0.185\textwidth, height=0.185\textwidth]{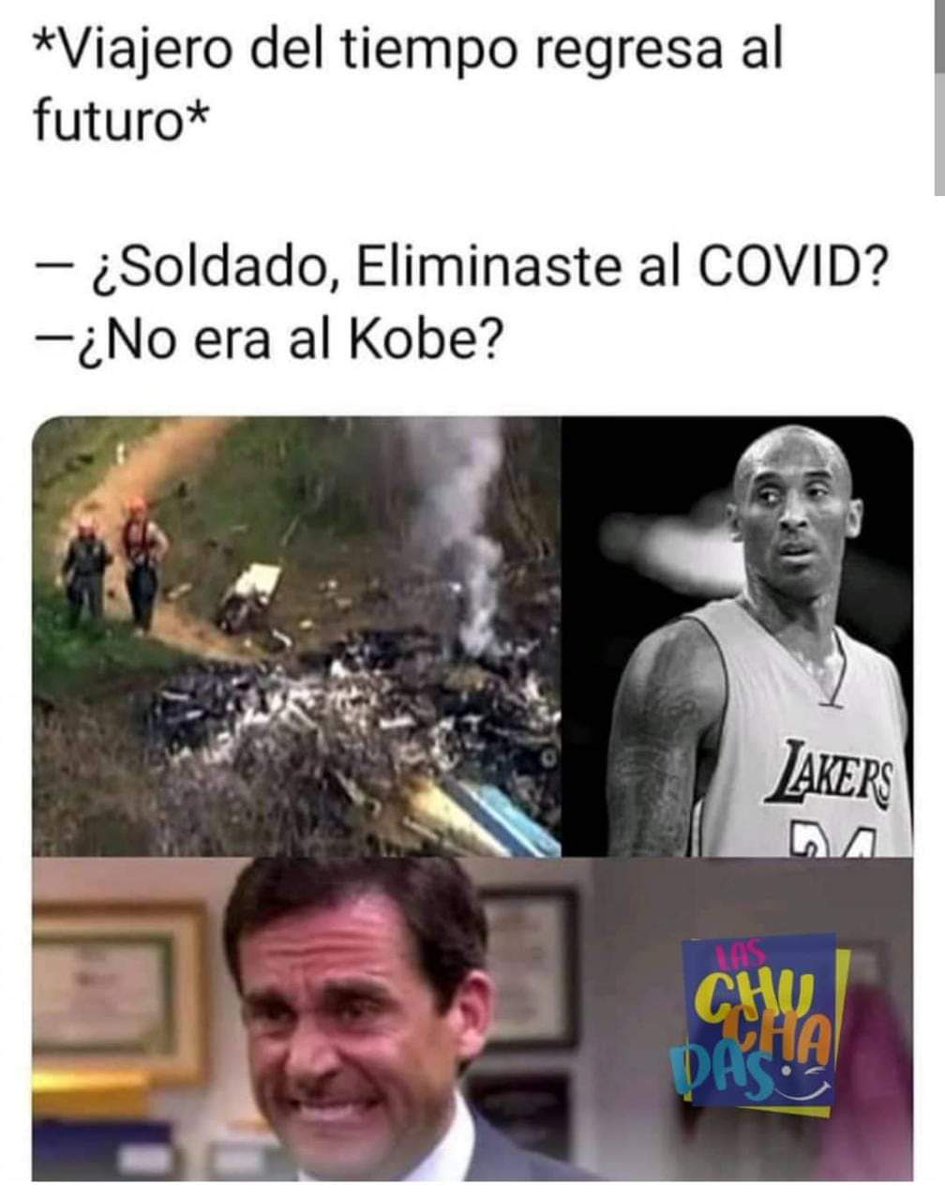}}\hspace{0.2em}
\subfloat[\label{fig:meme_rej:exm:2}]{
\includegraphics[width=0.185\textwidth, height=0.185\textwidth]{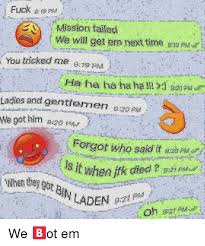}}\hspace{0.2em}
\subfloat[\label{fig:meme_rej:exm:3}]{
\includegraphics[width=0.185\textwidth, height=0.185\textwidth]{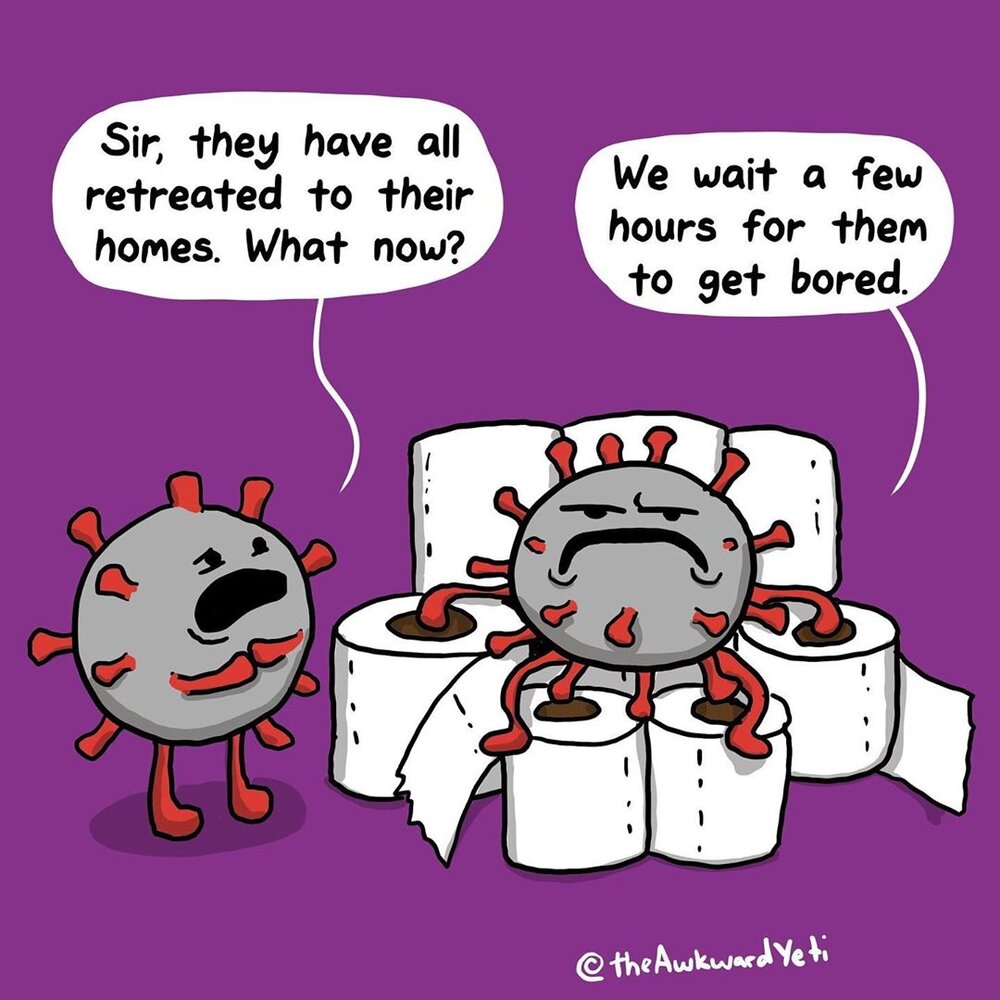}}\hspace{0.2em}
\subfloat[\label{fig:meme_rej:exm:4a}]{
\includegraphics[width=0.185\textwidth, height=0.185\textwidth]{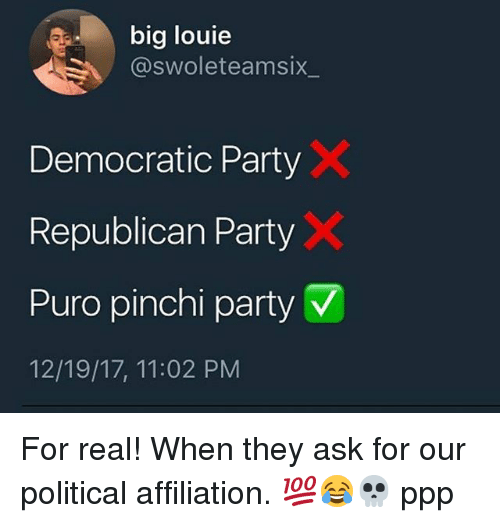}}\hspace{0.2em}
\subfloat[\label{fig:meme_rej:exm:4b}]{
\includegraphics[width=0.185\textwidth, height=0.185\textwidth]{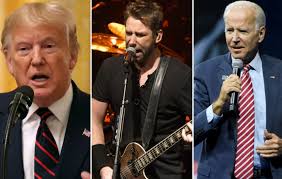}}\hspace{0.2em}
\caption{Examples of memes that we filtered out as part of the annotation process and the reason for that filtering: (a)~not in English, (b)~blurry text, (c)~cartoon, (d)~text-only, and (e)~image-only.}
\label{fig:meme:filtered_example}
\end{figure*}

Figure \ref{fig:meme:filtered_example} shows some example memes that we rejected during the filtering process due to them failing to satisfy some of the above criteria.

\section{Annotation Guidelines}\label{sec:annotation_guidelines}

\subsection{Defining \textit{harmful} memes}

The harm can be expressed in an obvious manner such as abusing, offending, disrespecting, insulting, demeaning, or disregarding a target entity or any socio-cultural or political ideology, belief, principle, or doctrine associated with that entity. The harm can also be in the form of a more subtle attack such as mocking or ridiculing a person or an idea.

Harmful memes can target a social entity (e.g.,~an individual, an organization, a community) and can aim at calumny/vilification/defamation based on their background (bias, social background, educational background, etc.). The harm can be in the form of mental abuse, psycho-physiological injury, proprietary damage, emotional disturbance, or public image damage. A harmful meme typically attacks celebrities or well-known organizations.

\subsection{The target categories}

Here are the categories for the targeted entities:
 
\begin{enumerate}  
   \item \textbf{Individual:} A person, usually a celebrity, e.g.,~a well-known politician, an actor, an artist, a scientist, an environmentalist, etc., such as \emph{Donald Trump, Joe Biden, Vladimir Putin, Hillary Clinton, Barack Obama, Chuck Norris, Greta Thunberg}, and \emph{Michelle Obama}.  
    \item \textbf{Organization:} A group of people with a particular purpose, such as a business, a government department, a company, an institution, or an association, e.g., \emph{Facebook}, \emph{WTO}, and the \emph{Democratic party}.

    \item \textbf{Community:} A social unit with commonalities based on personal, professional, social, cultural, or political attributes such as religious views, country of origin, gender identity, etc. Communities may share a sense of place situated in a given geographical area (e.g.,~a country, a village, a town, or a neighborhood) or in virtual space through communication platforms (e.g.,~online fora based on religion, country of origin, gender, etc.).
    \item \textbf{Society:} Society as a whole. When a meme promotes conspiracies or hate crimes, it becomes harmful to the general public, i.e.,~to the entire society. 
\end{enumerate}

\subsection{Characteristics of \textit{harmful} memes}

\begin{itemize}
    \item Harmful memes may or may not be offensive, hateful or biased in nature.
    \item Harmful memes point out vices, allegations, and other negative aspects of an entity based on verified or unfounded claims or mocks. 
    \item Harmful memes leave an open-ended connotation to the word \emph{community}, including antisocial communities such as terrorist groups.
    \item The harmful content in harmful memes is often implicit and might require critical judgment to establish the potency it can cause.
    \item Harmful memes can be classified on multiple levels, based on the intensity of the harm caused, e.g.,~\textit{very harmful}, \textit{partially harmful}. 
    \item A harmful meme can target multiple individuals, organizations, and/or communities at the same time. In such cases, we ask the annotators to go with their best personal choice. 
    \item Harm can take the form of sarcasm or satire. Sarcasm is praise that is actually an insult, and involves malice, the desire to demean someone. Satire is the ironical exposure of the vices or follies of an individual, a group, an institution, an idea, or society.

\end{itemize}

\begin{figure}[!t]
\begin{subfigure}[h]{.25\textwidth}
\centering
\includegraphics[width=1\textwidth]{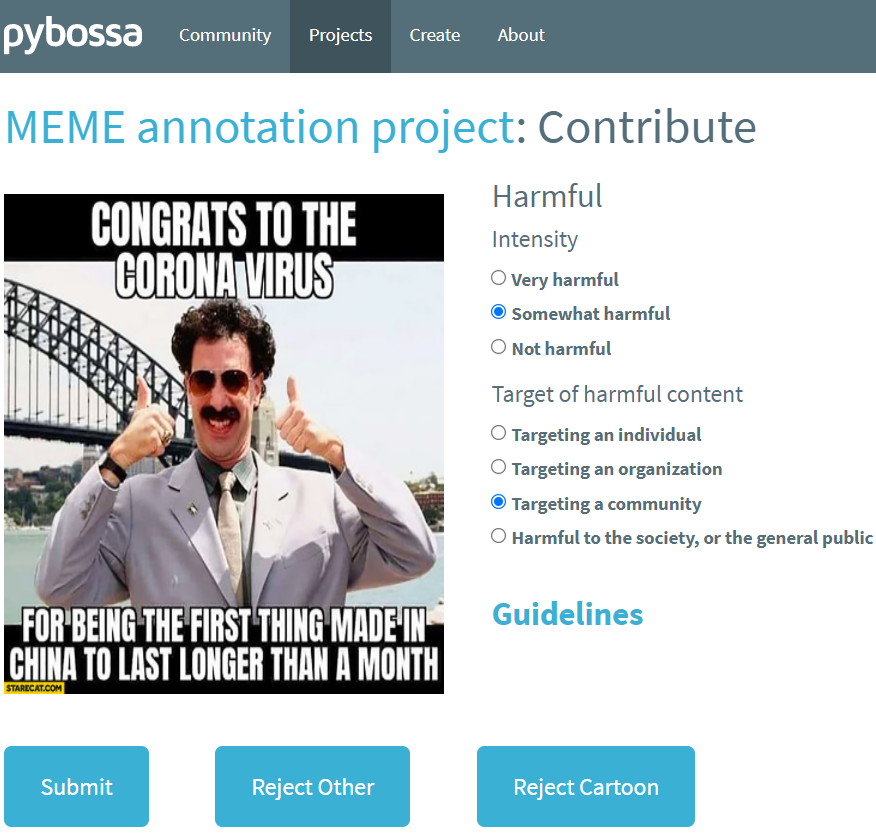}
\caption{Annotation interface}
\end{subfigure}%
\begin{subfigure}[h]{.25\textwidth}
\centering
\includegraphics[width=1\textwidth]{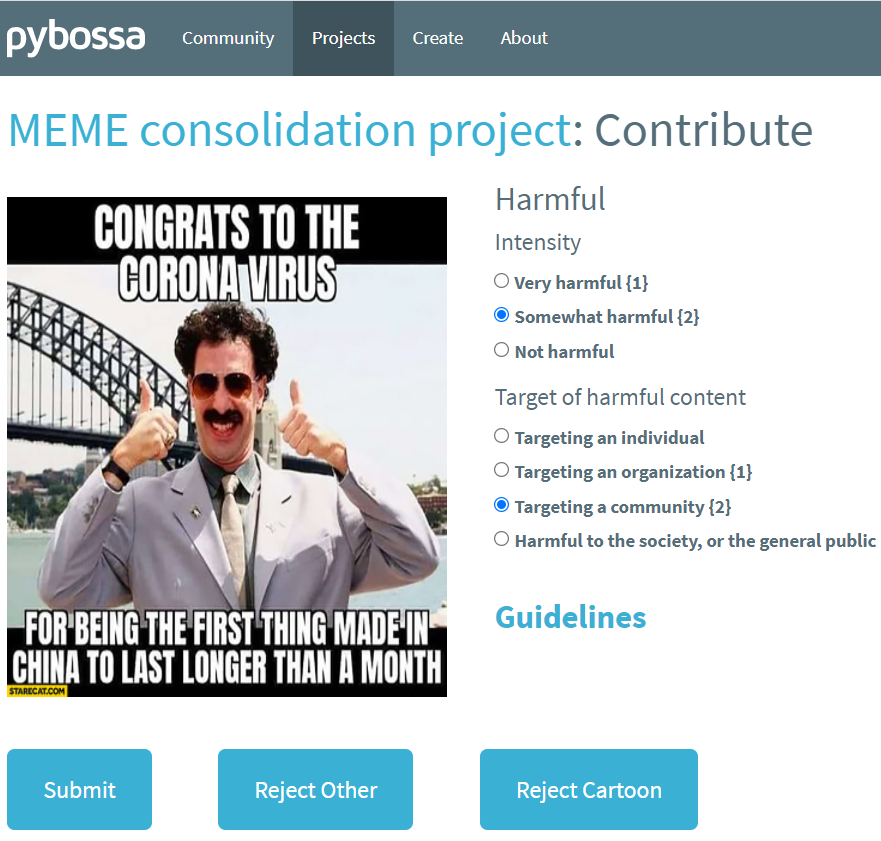}
\caption{Consolidation interface}
\end{subfigure}
\caption{Snapshot of the PyBossa GUI we used.}
\label{fig:platform}
\end{figure}

\begin{table*}[t!]
\centering
    
\resizebox{0.99\textwidth}{!}
{
\begin{tabular}{p{1.45cm} |l  l  l | l  l  l  l}
\toprule
\multirow{2}{1.45cm}{\bf Dataset}& \multicolumn{3}{c|}{\bf Harmfulness} & \multicolumn{4}{c}{\bf Target} \\ 
\cline{2-8}
& \em Very harmful & \em Partially harmful & \em Harmless & \em Individual & \em Organization & \em Community &  \em Society \\ 

\midrule

\multirow{5}{1.45cm}{\firstdata} & mask (\textcolor{black}{0.0512}) & trump (\textcolor{black}{0.0642}) & you (\textcolor{black}{0.0264}) & trump (\textcolor{black}{0.0541}) & deadline (\textcolor{black}{0.0709}) & china (\textcolor{black}{0.0665}) & mask (\textcolor{black}{0.0441}) \\

& trump (\textcolor{black}{0.0404}) &  president (\textcolor{black}{0.0273}) & home (\textcolor{black}{0.0263} & president (\textcolor{black}{0.0263}) & associated (\textcolor{black}{0.0709}) & chinese (\textcolor{black}{0.0417}) & vaccine	(\textcolor{black}{0.0430})\\

& wear (\textcolor{black}{0.0385}) & obama (\textcolor{black}{0.0262}) & corona (\textcolor{black}{0.0251}) & donald (\textcolor{black}{0.0231}) & extra (\textcolor{black}{0.0645}) & virus (\textcolor{black}{0.0361}) & alcohol (\textcolor{black}{0.0309})
\\

& thinks (\textcolor{black}{0.0308} & donald (\textcolor{black}{0.0241}) & work (\textcolor{black}{0.0222}) & obama (\textcolor{black}{0.0217}) & ensure (\textcolor{black}{0.0645}) & wuhan (\textcolor{black}{0.0359}) & temperatures (\textcolor{black}{0.0309}) \\

& killed (\textcolor{black}{0.0269}) & virus (\textcolor{black}{0.0213}) & day (\textcolor{black}{0.0188}) &
covid (\textcolor{black}{0.0203}) & qanon (\textcolor{black}{0.0600}) & cases (\textcolor{black}{0.0319}) & killed (\textcolor{black}{0.0271})\\

\midrule

\multirow{5}{1.45cm}{\seconddata} &  

photoshopped (\textcolor{black}{0.0589}) & democratic (\textcolor{black}{0.0164}) & party (\textcolor{black}{0.02514}) & biden (\textcolor{black}{0.0331}) & libertarian (\textcolor{black}{0.0358}) & liberals (\textcolor{black}{0.0328}) & crime (\textcolor{black}{0.0201}) \\
& married (\textcolor{black}{0.0343})& obama (\textcolor{black}{0.0158}) & debate (\textcolor{black}{0.0151}) & joe (\textcolor{black}{0.0323}) & republican (\textcolor{black}{0.0319}) &  radical (\textcolor{black}{0.0325}) & rights (\textcolor{black}{0.0195}) \\
& joe (\textcolor{black}{0.0309})& libertarian (\textcolor{black}{0.0156})  & president (\textcolor{black}{0.0139}) & obama (\textcolor{black}{0.0316}) & democratic (\textcolor{black}{0.0293}) & islam (\textcolor{black}{0.0323}) & gun (\textcolor{black}{0.0181}) \\
& trump (\textcolor{black}{0.0249}) & republican (\textcolor{black}{0.0140}) & democratic (\textcolor{black}{0.0111}) & trump  (\textcolor{black}{0.0286}) & green (\textcolor{black}{0.0146}) & black (\textcolor{black}{0.0237}) & taxes  (\textcolor{black}{0.0138}) \\
& nazis (\textcolor{black}{0.0241})& vote (\textcolor{black}{0.0096}) & green (\textcolor{black}{0.0086}) & putin (\textcolor{black}{0.0080}) & government (\textcolor{black}{0.0097}) & mexicans (\textcolor{black}{0.0168}) & law (\textcolor{black}{0.0135}) \\
\bottomrule
\end{tabular}}

\caption{The top-5 most frequent words in each class for each of the two tasks and each of the two datasets. The TF.IDF score for each word is shown in parentheses.}

\label{tab:textual_lexical}
\end{table*}

\subsection{Annotation Process} \label{appendix:annotationprocess}

Figure~\ref{fig:platform} shows the annotation and consolidation interface, based on the crowd-sourcing platform pybossa,\footnote{\url{http://pybossa.com/}} which we built for annotating degree of harmfulness and its target. Before starting the annotation process, we requested each annotator to thoroughly go through the annotation guidelines, and we conducted several discussion sessions to understand whether they understood what harmful content is and how to differentiate it from humorous, satirical, hateful, and non-harmful content. The average inter-annotator agreement scores in terms of Cohen's $\kappa$ for the two tasks are 0.683 and 0.782 on \firstdata, 0.675 and 0.790 on \seconddata.

\paragraph{Step 1. Training annotation.} First, we took a subset of 200 memes (100 per dataset), and we asked each annotator to do annotations for degree of harmfulness and for its target. This step aimed to ensure that annotators understood the definition of harmfulness and targets. After this initial step, the average inter-annotator agreement in terms of Cohen's $\kappa$ for the two tasks across all pairs of annotators was quite low:  0.241 and 0.317 on \firstdata, and 0.271 and 0.325 on \seconddata. Next, we asked the annotators to discuss their disagreements and to re-annotate the memes. This time, the average inter-annotator agreement in terms of Cohen's $\kappa$ improved to 0.704 and 0.815 on \firstdata, and to 0.711 and 0.826 on \seconddata, which is quite satisfactory for both tasks. Hence, we decided we were ready to start the actual annotation process. 

\paragraph{Step 2. Actual annotation.} In the actual annotation stage, we divided the two datasets into five equal subsets, and we assigned three annotators per subset. This ensured that each meme was annotated three times. We further asked the annotators to reject the memes that violated any of the filtering criteria, as described in Section~\ref{sec:filtering} above.

\paragraph{Step 3. Consolidation.} After the above annotation, the Cohen's $\kappa$ was quite good: it was 0.683 and 0.782 for \firstdata, and it was 0.675 and 0.790 for \seconddata. Yet, we observed many memes where two annotators chose the same label (e.g.,~\textit{partially harmful}), but the third one had made a different choice (e.g.,~\textit{very harmful}). To resolve such disagreements, in the consolidation phase, we used majority voting to decide on the final label. For cases where all three proposed labels were different, we involved an additional consolidator to help make the final decision. 

\begin{figure}[!t]
\begin{subfigure}[t]{.5\textwidth}
\centering
\includegraphics[width=1\textwidth]{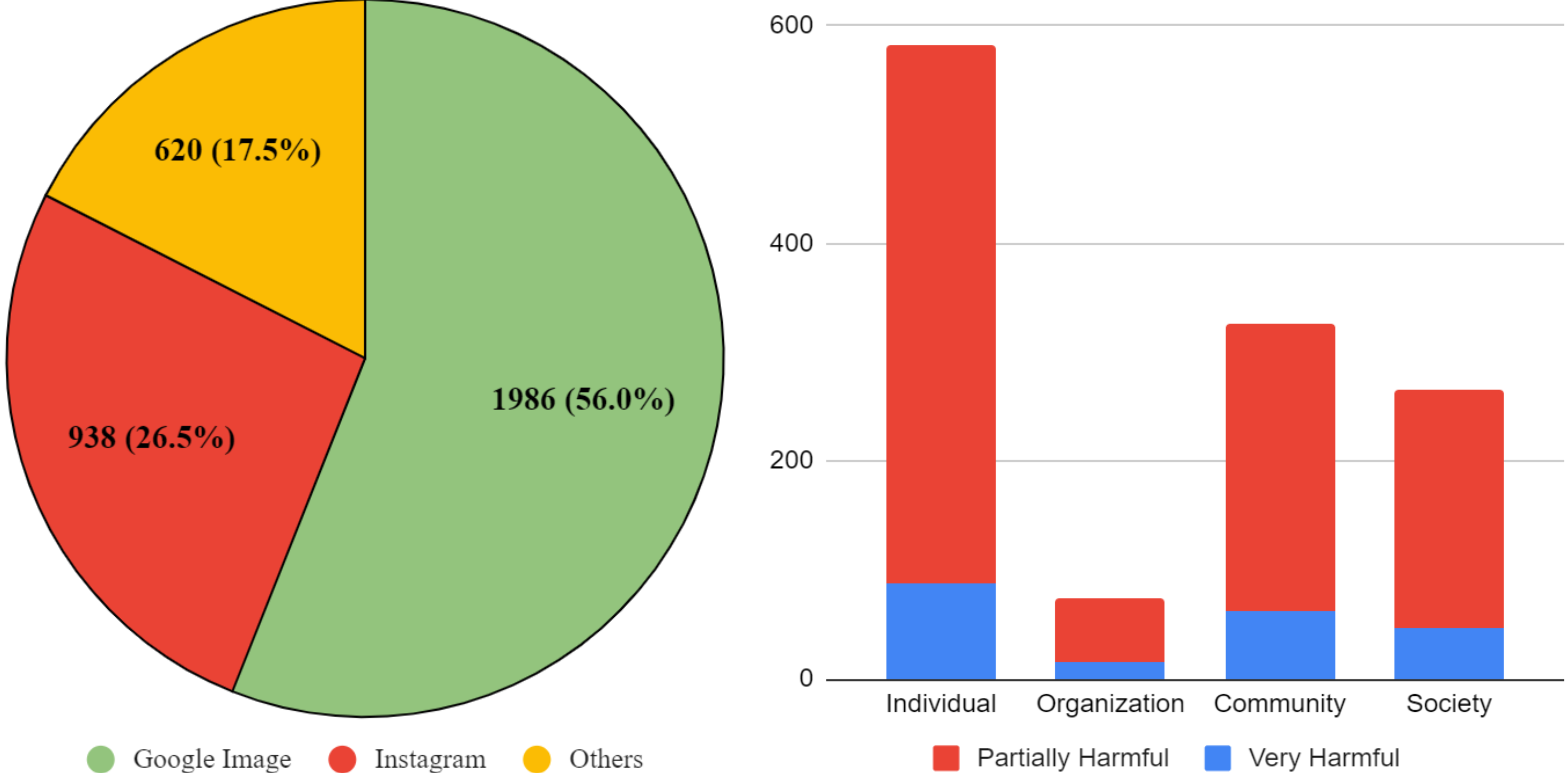}
\caption{statistics about \firstdata}
\end{subfigure}

\begin{subfigure}[b]{.5\textwidth}
\centering
\includegraphics[width=1\textwidth]{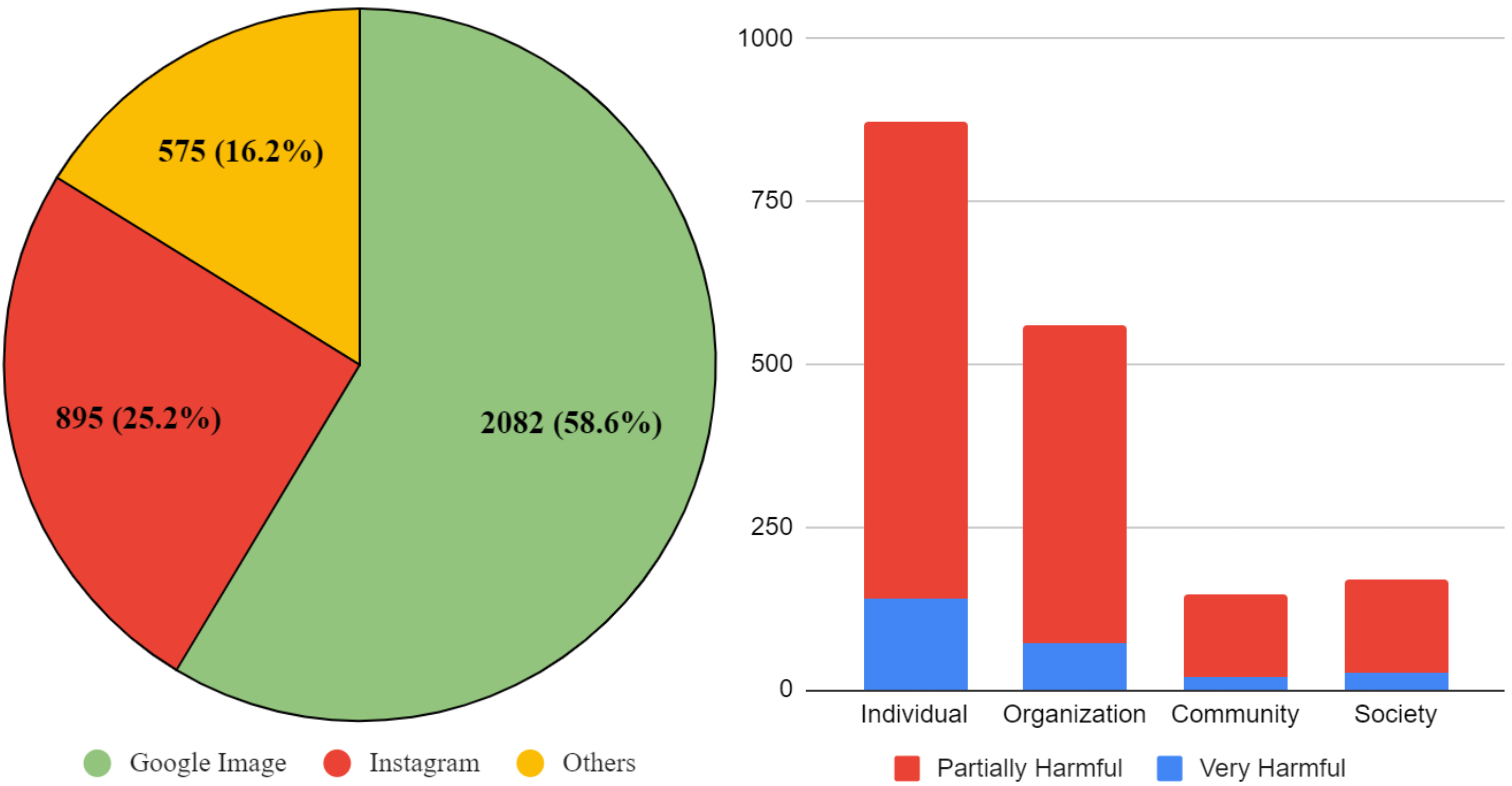}
\caption{statistics about \seconddata}
\end{subfigure}
\caption{Statistics about the distribution of sources and labels in the two datasets.}
\label{fig:source_label_statistics}
\end{figure}

Figure~\ref{fig:source_label_statistics} shows statistics about the distribution of sources and labels in the final datasets.

\begin{figure*}[h]
\begin{subfigure}[t]{\textwidth}
\centering
\includegraphics[width=1\textwidth]{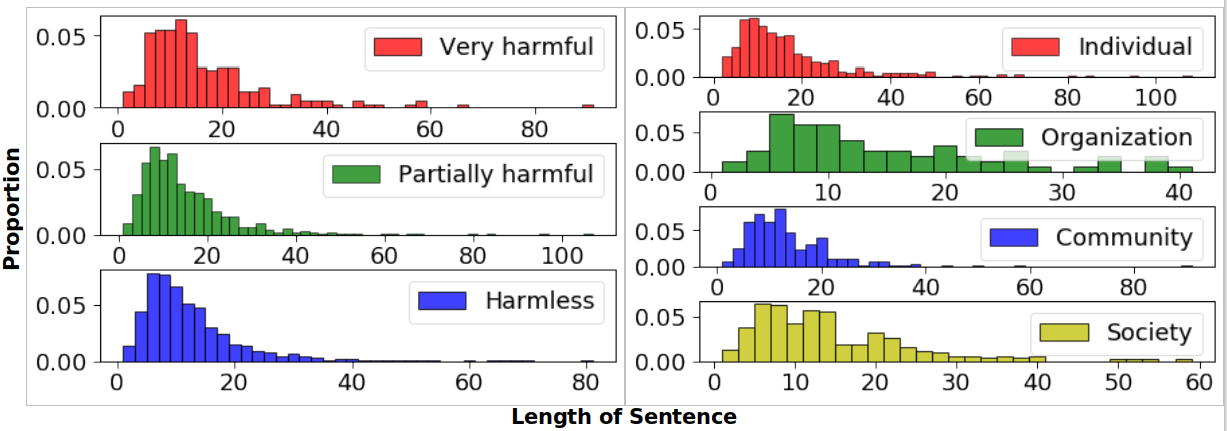}
\vspace{-5mm}
\caption{Meme text length distribution for \firstdata}
\end{subfigure}
\begin{subfigure}[t]{\textwidth}
\centering
\includegraphics[width=1\textwidth]{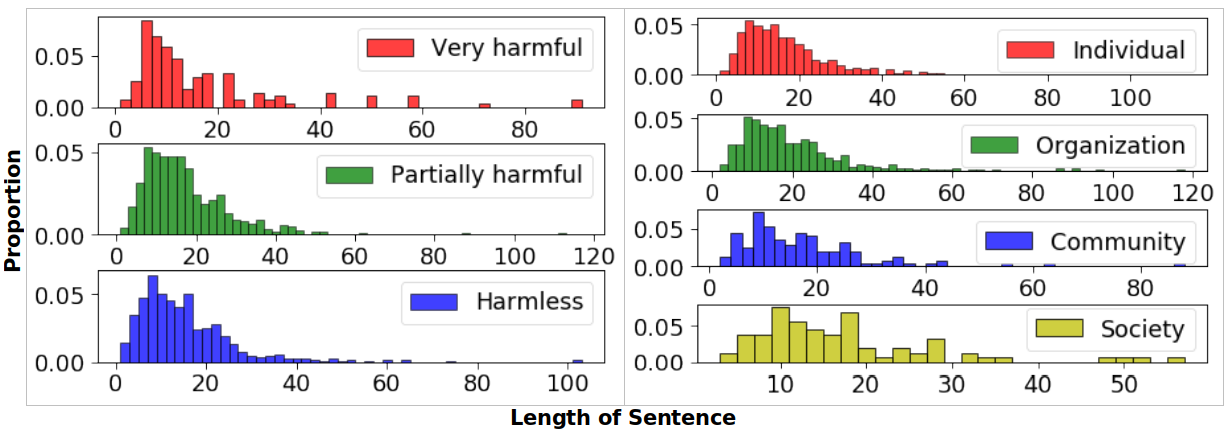}
\vspace{-5mm}
\caption{Meme text length distribution for \seconddata}
\end{subfigure}
\caption{Normalized histograms of meme text length per class for the two datasets.}
\label{fig:source_label}
\end{figure*}

\section{Lexical Statistics About the Datasets} \label{sec:lexical}

Table \ref{tab:textual_lexical} shows the top-5 most frequent words in the combined validation and test sets for the two datasets. We observe that for \textit{very harmful} and \textit{partially harmful} memes, the names of US politicians and COVID-19 oriented words are quite prominent. Moreover, we notice that the targets are dominated by words like \emph{Trump}, \emph{Joe}, \emph{Obama}, \emph{Republican}, \emph{Wuhan}, \emph{China}, \emph{Islam}, etc. To alleviate the potential bias caused by the presence of such words in the text of the memes, we intentionally included harmless memes related to these individuals, groups, and entities, as we have described in Section \ref{sec:dataset} above.

Figure \ref{fig:source_label} shows the length distribution of the meme text. We see that there are no major differences between the different classes.

\end{document}


\setlength\abovedisplayskip{2pt}
\setlength{\belowdisplayskip}{2pt}
\maketitle


\appendix


\section{Implementation Details and Hyperparameters} \label{hyperparameters}

We train all the models using Pytorch \citep{paszke2019pytorch} on a NVIDIA Tesla T4 GPU, with 16 GB dedicated memory, with CUDA-10 and cuDNN-11 installed. For the unimodal  models, we import all the pre-trained weights from TORCHVISION.MODELS\footnote{http://pytorch.org/docs/stable/torchvision/models.html} subpackage of the PyTorch framework. The non pre-trained weights are randomly initialized with a zero-mean Gaussian distribution with a standard deviation of 0.02. From the dataset statistics table (Table 1 in manuscript), we can observe a label imbalance problem for both harmfulness intensity ([\textit{Very Harmful}, \textit{Partially Harmful}] vs. \textit{Harmless}) and target ([\textit{Individual}, \textit{Organization}, \textit{Community}] vs. \textit{Entire Society}) classification tasks.
To deal with the imbalance problem, we use focal loss (FL) \cite{lin2017focal}, which down-weighs easy examples and focuses training on hard ones. We train \model\ in a multi-task learning setup, where the loss due to target identification is considered only if the meme is \textit{partially harmful} or \textit{very harmful}. We train our models using Adam \cite{kingma2014adam} optimizer and negative log-likelihood (NLL) loss as the objective function. In Table \ref{tab:hyperparameters}, we furnish the details of hyper-parameters used for the training.

\begin{table}[h]
\centering
\resizebox{0.99\columnwidth}{!}
{
\begin{tabular}{C | l | c | c | c | c | c | c }
& \multirow{2}{*}{\bf Models} & \multicolumn{6}{c}{\bf Hyperparameters} \\ \cline{3-8}
& &  \bf Batch-size & \bf Epochs & \bf Learning Rate & \bf Image Encoder & \bf Text Encoder & \bf \#Parameters \\ \hline

\multirow{5}{*}{\begin{turn}{-90} \centering \bf Unimodal \end{turn}} & TextBERT & 16 & 20 & 0.001 & - & Bert-base-uncased & 110,683,414\\
& VGG19 & 64 & 200 & 0.01 & VGG19 & - & 138,357,544\\
& DenseNet-161 & 32 & 200 & 0.01 & DenseNet-161 & - & 28,681,538\\
& ResNet-152 & 32 & 300 & 0.01 & ResNet-152 & - & 60,192,808\\
& ResNeXt-101 & 32 & 300 & 0.01 & ResNeXt-101 & - & 83,455,272\\
\hline 
\multirow{6}{*}{\begin{turn}{-90} \centering \bf Multimodal \end{turn}} & Late Fusion & 16 & 20 & 0.0001 & ResNet-152 & Bert-base-uncased & 170,983,752 \\
& Concat BERT & 16 & 20 & 0.001 & ResNet-152 & Bert-base-uncased & 170,982,214\\
& MMBT & 16 & 20 & 0.001 & ResNet-152 & Bert-base-uncased & 169,808,726\\
& ViLBERT CC & 16 & 10 & 0.001 & Faster RCNN & Bert-base-uncased & 112,044,290\\
& V-BERT COCO & 16 & 10 & 0.001 & Faster RCNN & Bert-base-uncased & 247,782,404\\ \cline{2-8}
& \model & 64 & 50 & 0.001 & VGG19 & DistilBERT-base-uncased & 7,608,323 \\

\hline
\end{tabular}}

\caption{Hyperparameters of different baseline models and \model.}

\label{tab:hyperparameters}
\vspace{-2mm}
\end{table}

\section{Filtering}\label{sec:filtering}

To assure a satisfactory quality of the \firstdata\ and \seconddata\ datasets, we impose four well-defined and fine-grained filtering criteria during the data collection and annotation process. The criteria are as follows:

\begin{enumerate}
    \item The meme text must not be in code-mixed or non-English language.
    \item The meme text must be readable. (e.g., blurry text, incomplete text, etc. are not allowed)
    \item The meme must not be unimodal, i.e. it should not contain only textual or visual content.
    \item The meme must not contain several cartoons. (we add these filtering criteria as cartoons are often very hard to be interpreted by AI systems)
\end{enumerate}

\begin{figure*}[ht!]
\centering
\subfloat[\label{fig:meme_rej:exm:1}]{
\includegraphics[width=0.185\textwidth, height=0.185\textwidth]{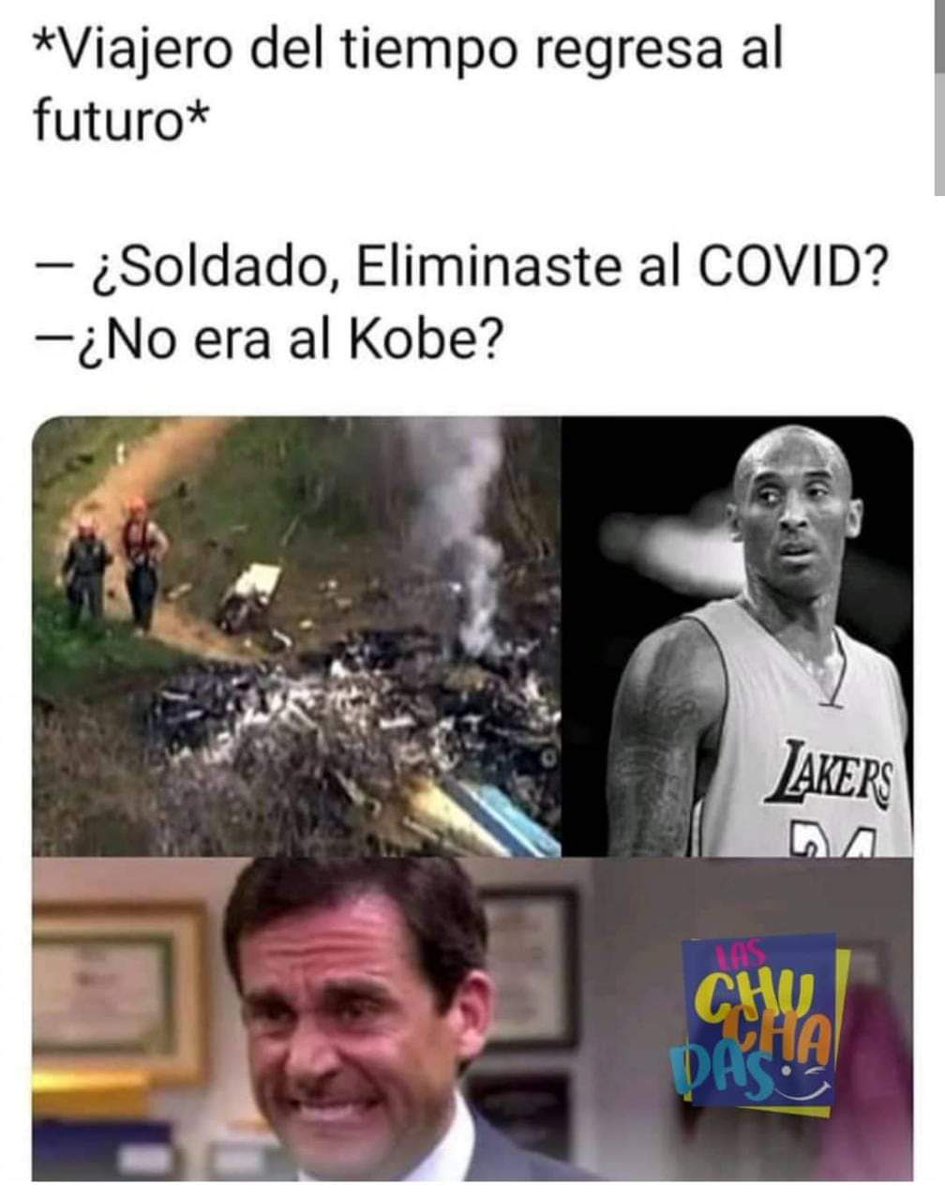}}\hspace{0.2em}
\subfloat[\label{fig:meme_rej:exm:2}]{
\includegraphics[width=0.185\textwidth, height=0.185\textwidth]{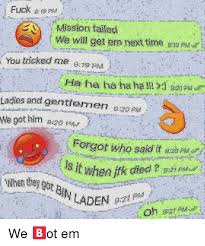}}\hspace{0.2em}
\subfloat[\label{fig:meme_rej:exm:3}]{
\includegraphics[width=0.185\textwidth, height=0.185\textwidth]{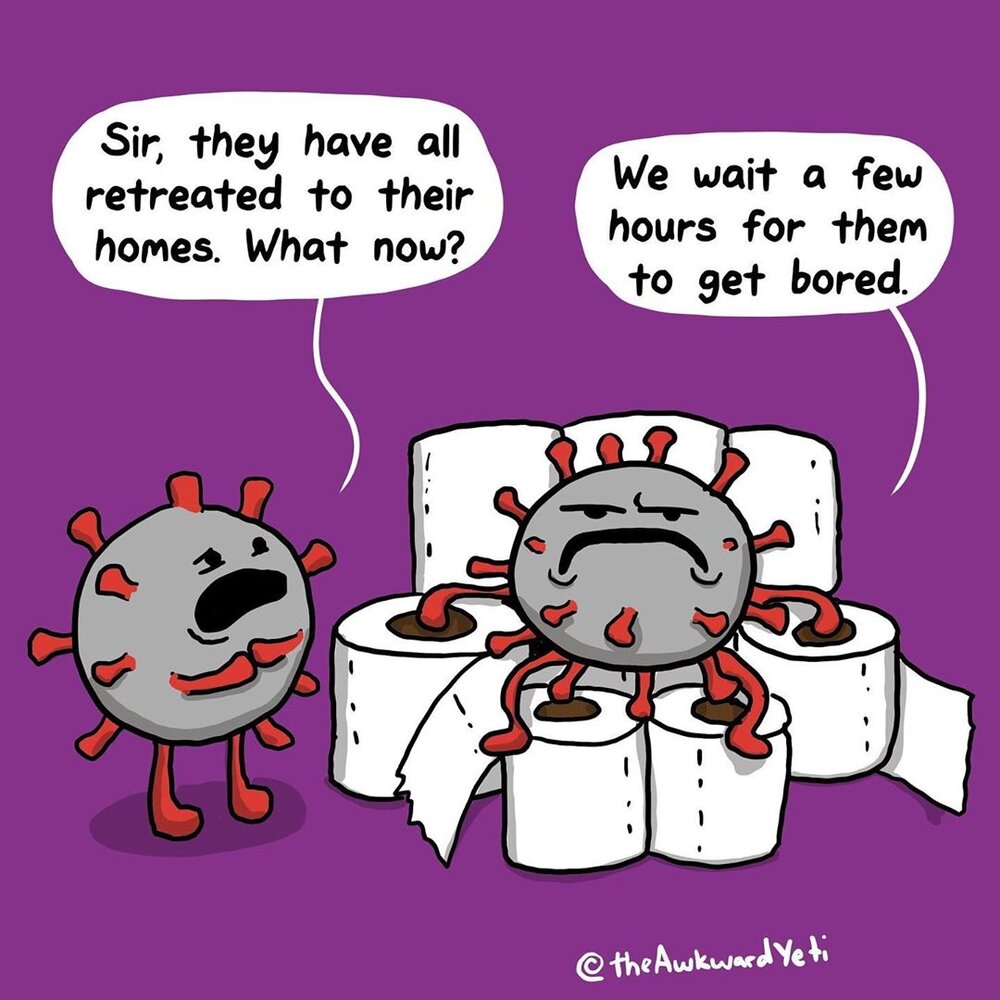}}\hspace{0.2em}
\subfloat[\label{fig:meme_rej:exm:4a}]{
\includegraphics[width=0.185\textwidth, height=0.185\textwidth]{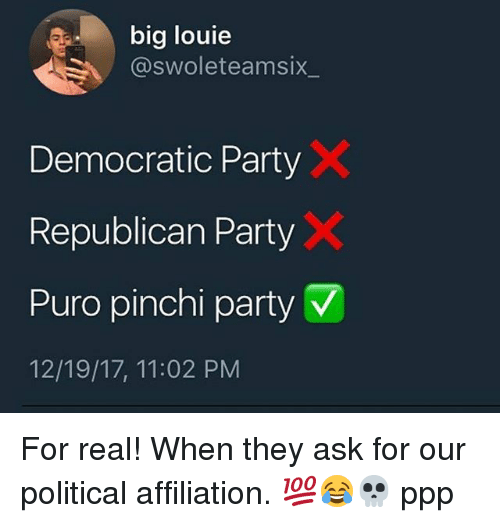}}\hspace{0.2em}
\subfloat[\label{fig:meme_rej:exm:4b}]{
\includegraphics[width=0.185\textwidth, height=0.185\textwidth]{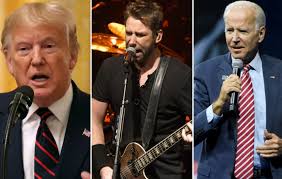}}\hspace{0.2em}
\caption{Example memes filtered out during annotation. Where meme sub-figures (a): Non-english, (b): Blurry text, (c): Cartoons, (d): Text-only (e): Image-only.}
\label{fig:meme:filtered_example}
\end{figure*}

Figure \ref{fig:meme:filtered_example} shows some example memes which were rejected during filtering process due to the four criteria mentioned above.

\section{Annotation Guidelines}\label{sec:annotation_guidelines}

\subsection{What do we mean by \textit{harmful} memes?}

The intended `harm' can be expressed in an obvious manner such as by abusing, offending, disrespecting, insulting, demeaning, or disregarding the entity or any socio-cultural or political ideology, belief, principle, or doctrine associated with that entity. Likewise, the `harm' can also be in the form of a more subtle attack such as by mocking or ridiculing a person or an idea.

The entrenched meaning of harmful memes is targeted towards a social entity (e.g., an individual, an organization, a community, etc.), likely to cause calumny/ vilification/ defamation depending on their background (bias, social background, educational background, etc.). The ‘harm’ caused by a meme can be in the form of mental abuse, psycho-physiological injury, proprietary damage, emotional disturbance, compensated public image. A harmful meme typically attacks celebrities or well-known organizations intending to expose their professional demeanor.

\subsection{What are the four target categories?}

 The four target entities are as follows:
 
\begin{enumerate}  
 [topsep=0pt,itemsep=-1ex,partopsep=1ex,parsep=1ex,leftmargin= 0.2in]
    \item \textbf{Individual:} A person, usually a celebrity (e.g., well-known politician, actor, artist, scientist, environmentalist, etc. such as Donald Trump, Joe Biden, Vladimir Putin, Hillary Clinton, Barack Obama, Chuck Norris, Greta Thunberg, Michelle Obama).  
    \item \textbf{Organization:} An organization is a group of people with a particular purpose, such as a business, government department, company, institution or association – comprising more than one person, and having a particular purpose, such as research organizations (e.g., WTO, Google) and political organizations (e.g., Democratic party).

    \item \textbf{Community:} A community is a social unit with commonalities based on personal, professional, social, cultural, or political attributes such as religious views, country of origin, gender identity, etc. Communities may share a sense of place situated in a given geographical area (e.g., a country, village, town, or neighborhood) or in virtual space through communication platforms (e.g., online forums based on religion, country of origin, gender).
    \item \textbf{Society:} When a meme promotes conspiracies or hate crimes, it becomes harmful to general public, i.e., the entire society. 
\end{enumerate}

\begin{table*}[t!]
\centering
    
\resizebox{0.99\textwidth}{!}
{
\begin{tabular}{p{1.45cm} |c | c | c || c | c | c | c}
\multirow{2}{1.45cm}{\bf Dataset}& \multicolumn{3}{c||}{Harmfulness} & \multicolumn{4}{c}{Target} \\ 
\cline{2-8}
& Very harmful & Partially harmful & Harmless & Individual & Organization & Community &  Society \\ 

\hline

\multirow{5}{1.45cm}{\firstdata} & mask (\textcolor{black}{0.0512}) & trump (\textcolor{black}{0.0642}) & you (\textcolor{black}{0.0264}) & trump (\textcolor{black}{0.0541}) & deadline (\textcolor{black}{0.0709}) & china (\textcolor{black}{0.0665}) & mask (\textcolor{black}{0.0441}) \\

& trump (\textcolor{black}{0.0404}) &  president (\textcolor{black}{0.0273}) & home (\textcolor{black}{0.0263} & president (\textcolor{black}{0.0263}) & associated (\textcolor{black}{0.0709}) & chinese (\textcolor{black}{0.0417}) & vaccine	(\textcolor{black}{0.0430})\\

& wear (\textcolor{black}{0.0385}) & obama (\textcolor{black}{0.0262}) & corona (\textcolor{black}{0.0251}) & donald (\textcolor{black}{0.0231}) & extra (\textcolor{black}{0.0645}) & virus (\textcolor{black}{0.0361}) & alcohol (\textcolor{black}{0.0309})
\\

& thinks (\textcolor{black}{0.0308} & donald (\textcolor{black}{0.0241}) & work (\textcolor{black}{0.0222}) & obama (\textcolor{black}{0.0217}) & ensure (\textcolor{black}{0.0645}) & wuhan (\textcolor{black}{0.0359}) & temperatures (\textcolor{black}{0.0309}) \\

& killed (\textcolor{black}{0.0269}) & virus (\textcolor{black}{0.0213}) & day (\textcolor{black}{0.0188}) &
covid (\textcolor{black}{0.0203}) & qanon (\textcolor{black}{0.0600}) & cases (\textcolor{black}{0.0319}) & killed (\textcolor{black}{0.0271})\\

\hline
\hline

\multirow{5}{1.45cm}{\seconddata} &  

photoshopped (\textcolor{black}{0.0589}) & democratic (\textcolor{black}{0.0164}) & party (\textcolor{black}{0.02514}) & biden (\textcolor{black}{0.0331}) & libertarian (\textcolor{black}{0.0358}) & liberals (\textcolor{black}{0.0328}) & crime (\textcolor{black}{0.0201}) \\
& married (\textcolor{black}{0.0343})& obama (\textcolor{black}{0.0158}) & debate (\textcolor{black}{0.0151}) & joe (\textcolor{black}{0.0323}) & republican (\textcolor{black}{0.0319}) &  radical (\textcolor{black}{0.0325}) & rights (\textcolor{black}{0.0195}) \\
& joe (\textcolor{black}{0.0309})& libertarian (\textcolor{black}{0.0156})  & president (\textcolor{black}{0.0139}) & obama (\textcolor{black}{0.0316}) & democratic (\textcolor{black}{0.0293}) & islam (\textcolor{black}{0.0323}) & gun (\textcolor{black}{0.0181}) \\
& trump (\textcolor{black}{0.0249}) & republican (\textcolor{black}{0.0140}) & democratic (\textcolor{black}{0.0111}) & trump  (\textcolor{black}{0.0286}) & green (\textcolor{black}{0.0146}) & black (\textcolor{black}{0.0237}) & taxes  (\textcolor{black}{0.0138}) \\
& nazis (\textcolor{black}{0.0241})& vote (\textcolor{black}{0.0096}) & green (\textcolor{black}{0.0086}) & putin (\textcolor{black}{0.0080}) & government (\textcolor{black}{0.0097}) & mexicans (\textcolor{black}{0.0168}) & law (\textcolor{black}{0.0135}) \\
\hline
\end{tabular}}

\caption{Top-5 most frequent words per class for the \firstdata\ and \seconddata\ datasets. The tf-idf score per word  is given within parenthesis.}

\label{tab:textual_lexical}
\end{table*}

\subsection{Characteristics of \textit{harmful} memes:}

\begin{itemize}
    \item Harmful memes may or may not be offensive, hateful or biased in nature.
    \item Harmful memes point out vices, allegations, and other negative aspects of an entity based on verified or unfounded claims or mocks. 
    \item Harmful memes leave an open-ended connotation to the word ‘community’, including ‘antisocial’ communities such as terrorist groups.
    \item The harmful content in harmful memes is often implicit and might require critical judgment to establish the potency it can cause.
    \item Harmful memes can be classified on multiple levels, based on the intensity of the harm caused, e.g., \textit{very harmful}, \textit{partially harmful}. 
    \item One harmful meme can target multiple individual, organizations, communities at the same time. In that case, we asked the annotators to go with the best personal judgment. 
    \item Harm can be expressed in form of sarcasm and/or political satire. Sarcasm is praise which is really an insult; sarcasm generally involves malice, the desire to put someone down. On the other hand, satire is the ironical exposure of the vices or follies of an individual, a group, an institution, an idea, a society, etc., usually with a view to correcting it.

\end{itemize}

\begin{figure}[!t]
\begin{subfigure}[h]{.25\textwidth}
\centering
\includegraphics[width=1\textwidth]{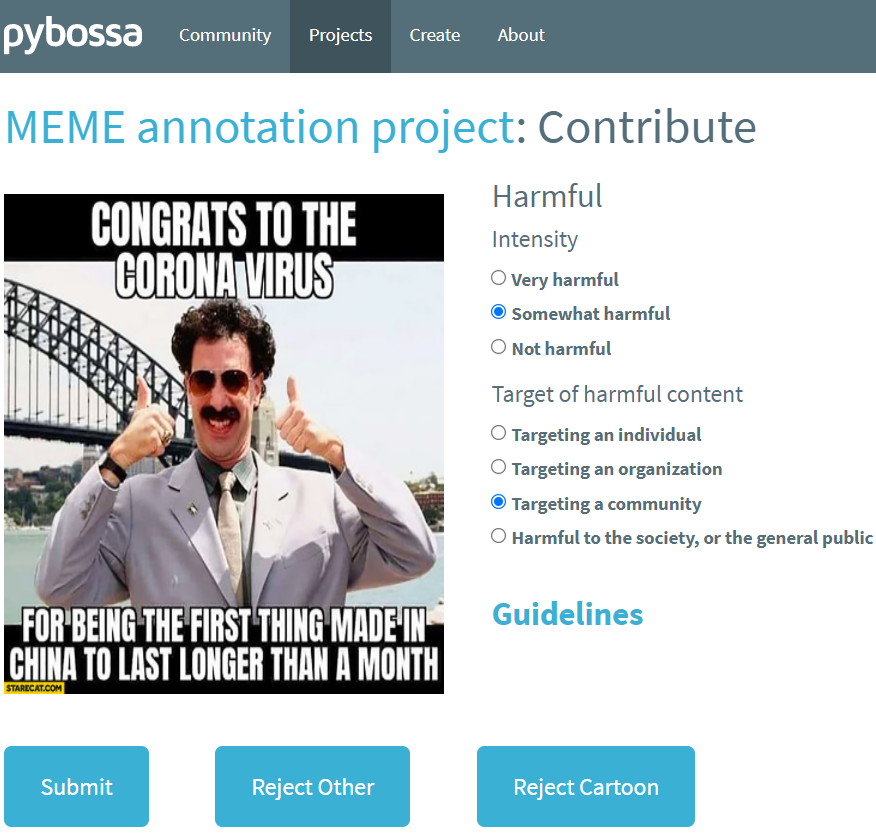}
\caption{Annotation interface}
\label{fig:annotation}
\end{subfigure}%
\begin{subfigure}[h]{.25\textwidth}
\centering
\includegraphics[width=1\textwidth]{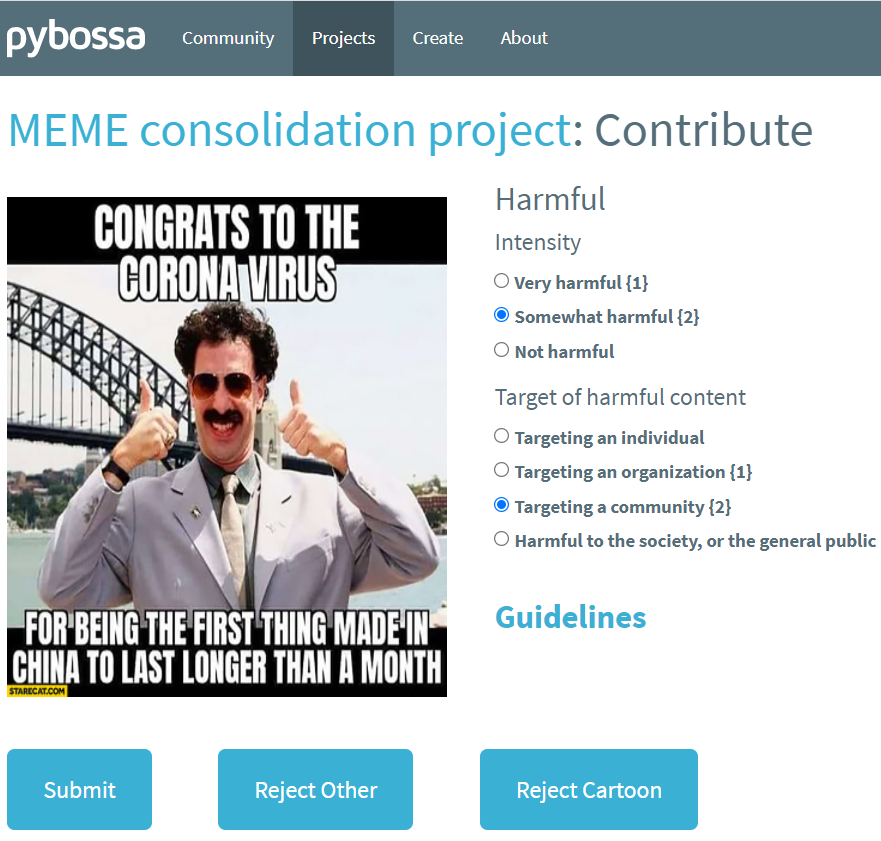}
\caption{Consolidation interface}
\label{fig:consolidation}
\end{subfigure}
\caption{Snapshot of the PyBossa GUI  used for annotation and consolidation.}
\label{fig:platform}
\end{figure}

\section{Annotation Process}\label{appendix:annotationprocess}
We use the crowd-sourcing platform pybossa\footnote{\url{https://pybossa.com/}}
(c.f. Figure \ref{fig:platform}) to build an annotation interface that shows each meme and request annotations for harmfulness levels and target. 
Before beginning the annotation process, we requested every annotator to thoroughly go through the annotation guidelines and conducted several discussion sessions to understand if all of them can understand exactly what harmful content is and how to differentiate it from humorous, satirical, hateful, and non-harmful content. The average inter-annotator agreement scores (Cohen's $\kappa$) \cite{bobicev-sokolova-2017-inter} for \firstdata\ and \seconddata\ $0.683$ and $0.790$, respectively.

\noindent\textbf{Step 1 - Dry run:} At first, we took a subset of 200 memes ($100$ from each dataset) and requested each annotator to annotate for harmfulness levels and targets. This step aimed to ensure that each annotator can comprehend the definition of harmfulness and targets. After this step, the average inter-annotator agreement score (Cohen's $\kappa$) \cite{bobicev-sokolova-2017-inter} for two tasks across all pairs of annotators was only $0.241$ and $0.325$, which is low but expected. The annotators discussed their disagreements and re-annotated the memes. This time, the score improved to $0.704$ and $0.826$, which is satisfactory for both tasks. Hence, we decided to begin the final annotation process. 

\noindent\textbf{Step 2 - Final annotation:} In the final annotation stage, we divided the two datasets into $5$ equal subsets and assigned $3$ annotators for each subset. This ensures that each meme is annotated $3$ times. We also request the annotators to reject the memes which violets filtering criteria, as described in Appendix \ref{sec:filtering}. This process engages an additional level of filtering to ensure adequate quality of the datasets. 

\noindent\textbf{Step 3 - Consolidation:} After the final annotation, the average inter-annotator agreement score over two whole datasets are $0.683$ and $0.790$. We observed many memes where the annotation of two annotators differs from the third one - for example, two annotators independently annotated a meme to be \textit{partially harmful}, but the third annotator annotated as \textit{very harmful}. In the consolidation phase, we used majority voting to decide the final label. For cases where all the three annotations are different, we employed a fourth annotator to take the final decision. 

\begin{figure}[!t]
\begin{subfigure}[t]{.5\textwidth}
\centering
\includegraphics[width=1\textwidth]{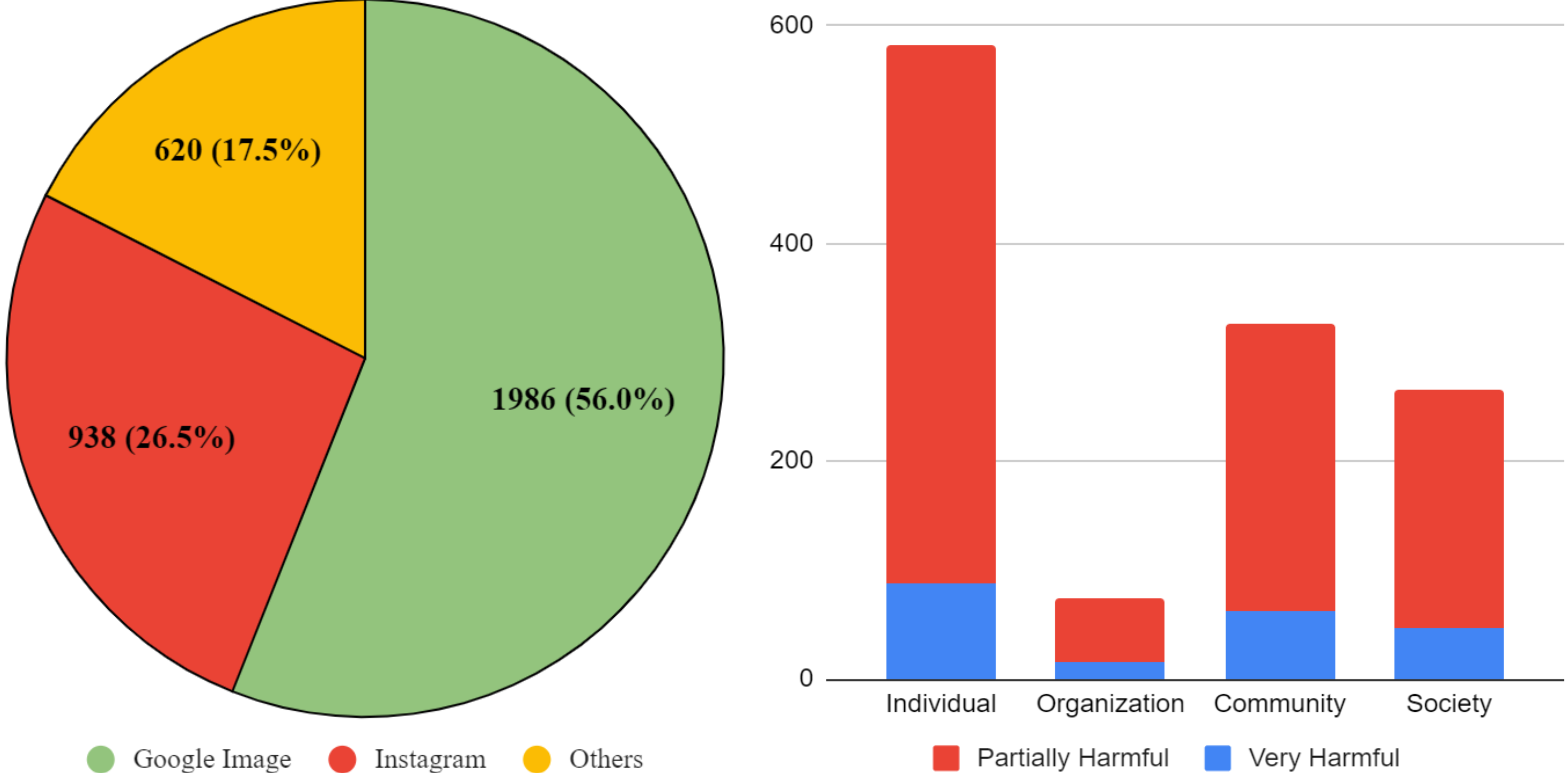}
\caption{Source and label statistics of \firstdata\ dataset.}
\label{fig:annotation}
\end{subfigure}

\begin{subfigure}[b]{.5\textwidth}
\centering
\includegraphics[width=1\textwidth]{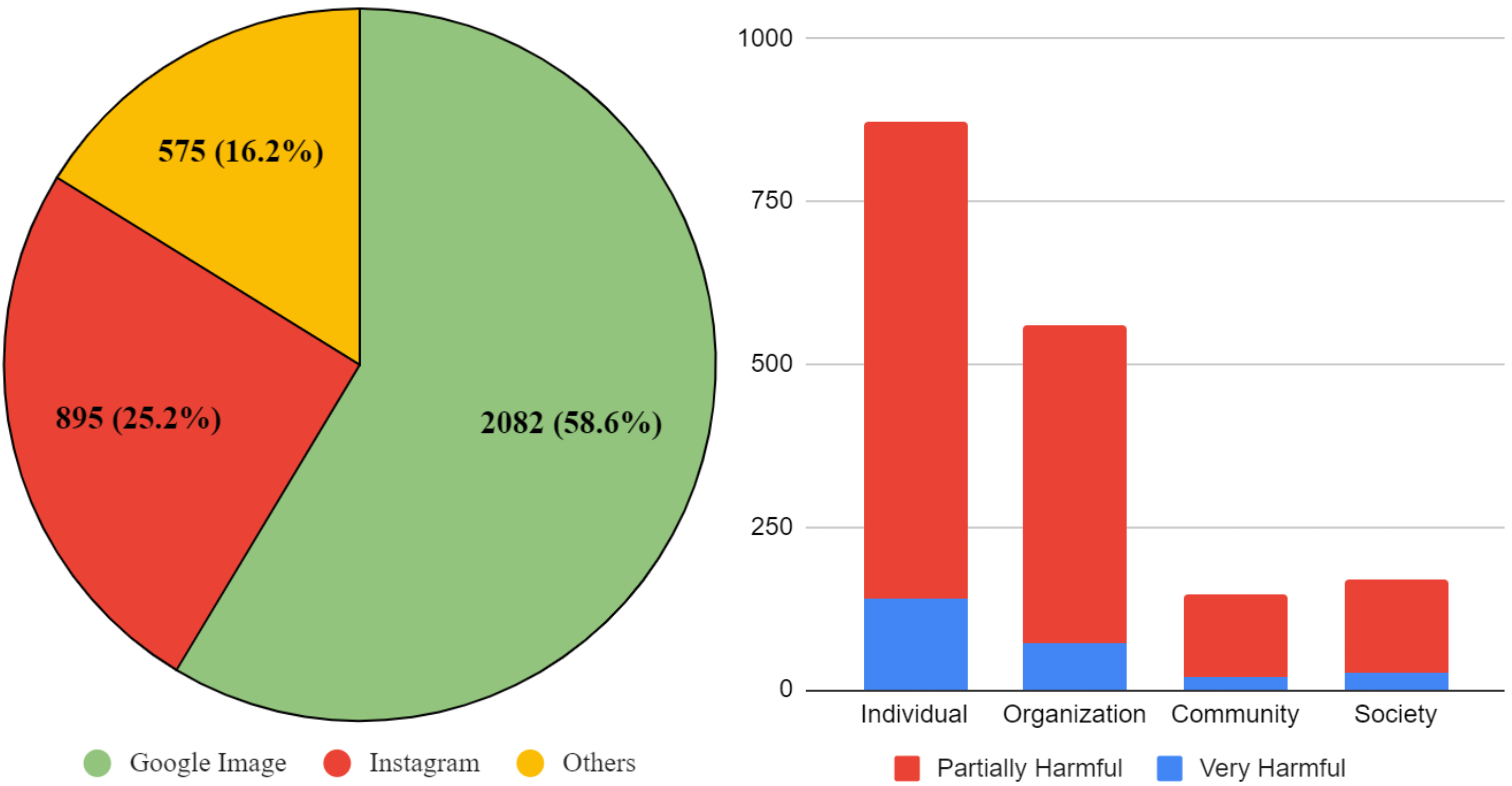}
\caption{Source and label statistics of \seconddata\ dataset.}
\label{fig:consolidation}
\end{subfigure}
\caption{Source and label statistics of two datasets.}
\label{fig:source_label}
\end{figure}

\section{Lexical Statistics of Two Datasets} \label{sec:lexical}

This section analyzes the lexical (word-level) statistics of the \firstdata\ and \seconddata\ datasets. Figure \ref{fig:source_label} shows the length distribution of the meme text for both the tasks across two datasets. Furthermore, Table \ref{tab:textual_lexical} shows the top-5 most frequent words by every separate class in the combined validation and test sets of two datasets. We observe that for \textit{very harmful} and \textit{partially harmful} classes, names of US politicians and COVID-19 oriented words are frequent. In the target classes, we notice the presence of various class-specific words such as `trump', 'joe', `obama', 'republican', 'wuhan', `china', 'islam'. To alleviate potential bias caused by these class-specific systems, we intentionally included harmless memes of related to these individuals, groups and entities, which is described more in the Section $4$ of main manuscript.

\begin{figure}[h]
\begin{subfigure}[t]{.5\textwidth}
\centering
\includegraphics[width=1\textwidth]{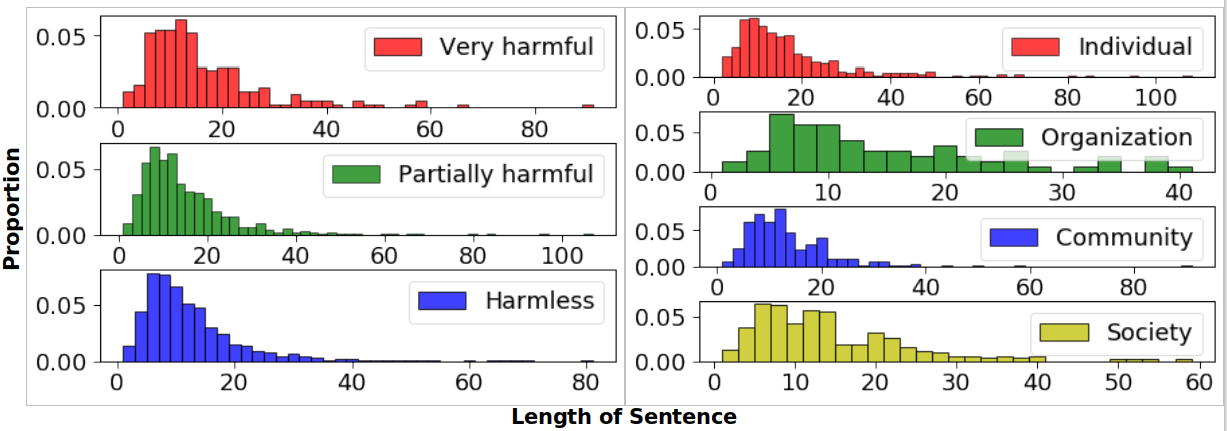}
\label{fig:annotation}
\vspace{-5mm}
\caption{Meme text length distribution for \firstdata.}
\end{subfigure}
\begin{subfigure}[t]{.5\textwidth}
\centering
\includegraphics[width=1\textwidth]{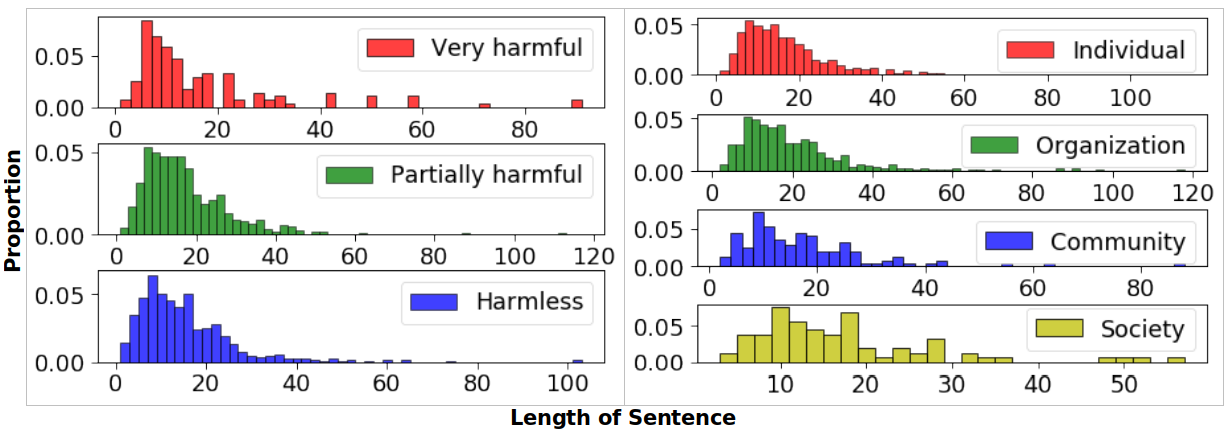}
\label{fig:consolidation}
\vspace{-5mm}
\caption{Meme text length distribution for \seconddata.}
\end{subfigure}
\caption{Normalized histograms of meme text length per class for two datasets.}
\label{fig:source_label}
\end{figure}

\bibliography{anthology,custom}
\bibliographystyle{acl_natbib}